\documentclass[12pt]{article} 
\usepackage[sectionbib]{natbib}
\usepackage{array,epsfig,fancyheadings,rotating}
\usepackage[]{hyperref}  
\usepackage{sectsty, secdot}
\sectionfont{\fontsize{12}{14pt plus.8pt minus .6pt}\selectfont}
\renewcommand{\theequation}{\thesection\arabic{equation}}
\subsectionfont{\fontsize{12}{14pt plus.8pt minus .6pt}\selectfont}

\usepackage{amsmath}
\usepackage{amssymb}
\usepackage{amsfonts}
\usepackage{multirow}
\usepackage{amsthm}

\newtheorem{theorem}{Theorem}

\newtheorem{corollary}{Corollary}
\newtheorem{proposition}{Proposition}
\theoremstyle{definition}

\newtheorem{assumption}{Assumption}

\usepackage{float}
\usepackage{mathrsfs}
\usepackage{booktabs}
\usepackage{rotating}
\usepackage{color}
\usepackage{caption}
\usepackage{rotating}
\usepackage{graphicx}
\usepackage{mathtools}
\usepackage{url}
\usepackage{pdflscape}
\usepackage{subfigure}
\usepackage{epstopdf}
\usepackage{bm}
\def\lsk{\left(}
\def\rsk{\right)}
\def\lbk{\left \{ }
\def\rbk{\right \} }
\def\lmk{\left [ }
\def\rmk{\right ] }
\DeclareMathOperator*{\argmin}{argmin}
\DeclareMathOperator*{\esssup}{ess\,sup}
\DeclareMathOperator*{\essinf}{ess\,inf}
\newcommand{\defeq}{\vcentcolon=}
\newcommand{\vect}[1]{\boldsymbol{#1}}
\newcommand{\norm}[1]{\left\Vert#1\right\Vert}

\newcommand{\be}{\vect{e}}

\newcommand{\bp}{\vect{p}}

\newcommand{\by}{\vect{y}}

\newcommand{\bv}{\vect{v}}

\newcommand{\bB}{\vect{B}}

\newcommand{\bE}{\vect{E}}
\newcommand{\bX}{\vect{X}}
\newcommand{\bY}{\vect{Y}}

\newcommand{\bZ}{\vect{Z}}

\newcommand{\bbeta}{\boldsymbol{\beta}}

\newcommand{\bgamma}{\boldsymbol{\gamma}}

\newcommand{\bGamma}{\boldsymbol{\Gamma}}
\newcommand{\bPhi}{\boldsymbol{\Phi}}

\newcommand{\bzero}{\mathbf{0}}
\newcommand{\bone}{\mathbf{1}}
\newcommand{\tv}{\text{vec}}

\newcommand{\hgamma}{\hat{\bgamma}}
\newcommand{\hGamma}{\hat{\bGamma}}
\newcommand{\hbeta}{\hat{\bbeta}}
\newcommand{\bQ}{\mathbb{Q}}
\newcommand{\tPhi}{\tilde{\Phi}}

\newcommand{\tB}{\tilde{\bB}}
\newcommand{\tX}{\tilde{\bX}}

\newcommand{\bbetaJC}{\bbeta^*_{J_{\eta}^C}}
\begin{document}
	\renewcommand{\baselinestretch}{2}
	\markright{ \hbox{\footnotesize\rm Statistica Sinica
		}\hfill\\[-13pt]
		\hbox{\footnotesize\rm
		}\hfill }
	
	\markboth{\hfill{\footnotesize\rm Zhenzhong Wang, Abolfazl Safikhani, Zhengyuan Zhu and David S.\ Matteson} \hfill}
	{\hfill {\footnotesize\rm High-dimensional Spatio-temporal VAR} \hfill}
	
	\renewcommand{\thefootnote}{}
	$\ $\par
	
	
	\fontsize{12}{14pt plus.8pt minus .6pt}\selectfont \vspace{0.8pc}
	\centerline{\large\bf Regularized Estimation in High-Dimensional Vector}
	\vspace{2pt} \centerline{\large\bf Auto-Regressive Models using Spatio-Temporal Information}
	\vspace{.4cm} \centerline{Zhenzhong Wang, Abolfazl Safikhani, Zhengyuan Zhu and David S.\ Matteson} \vspace{.4cm} \centerline{\it
		Iowa State University, University of Florida and Cornell University} \vspace{.55cm} \fontsize{9}{11.5pt plus.8pt minus
		.6pt}\selectfont
	
	
	\begin{quotation}
		\noindent {\it Abstract:}
A Vector Auto-Regressive (VAR) model is commonly used to model multivariate time series, and there are many penalized methods to handle high dimensionality. However in terms of spatio-temporal data, most methods do not take the spatial and temporal structure of the data into consideration, which may lead to unreliable network detection and inaccurate forecasts. This paper proposes a data-driven weighted $l_1$ regularized approach for spatio-temporal VAR model. 
		Extensive simulation studies are carried out to compare the proposed method with four existing methods of high-dimensional VAR model, demonstrating improvements of our method over others in parameter estimation, network detection and out-of-sample forecasts. We also apply our method on a traffic data set to evaluate its performance in real  application. In addition, we explore the theoretical properties of $l_1$ regularized estimation of VAR model under the weakly sparse scenario, in which the exact sparsity can be viewed as a special case. To the best of our knowledge, this direction has not been considered yet in the literature. 
For general stationary VAR process, we derive the non-asymptotic upper bounds on $l_1$ regularized estimation errors under the weakly sparse scenario, provide the conditions of estimation consistency, and further simplify these conditions for a special VAR(1) case. 

		\vspace{9pt}
		\par
	\end{quotation}\par
	\def\thefigure{\arabic{figure}}
	\def\thetable{\arabic{table}}
	\renewcommand{\theequation}{\thesection.\arabic{equation}}
	\fontsize{12}{14pt plus.8pt minus .6pt}\selectfont
	\setcounter{section}{1} 
	\setcounter{equation}{0} 
	\lhead[\footnotesize\thepage\fancyplain{}\leftmark]{}\rhead[]{\fancyplain{}\rightmark\footnotesize\thepage}
	
\noindent {\bf 1. Introduction}
\label{sec:intro}

The Vector Auto-regressive (VAR) model, a popular tool to simultaneously model and forecast a number of time series, has been widely applied in different scientific fields such as econometrics \citep{Sims1980}, finance \citep{Tsay2015}, ecology \citep{Hamptonetal} and so on. Recent developments in computing have made high-dimensional time series increasingly common in many studies. As the number of time series component increases, the number of parameters in VAR model increases dramatically, which leads to unreliable or even infeasible estimation. The usual way to handle the high dimensionality is to impose sparsity or low rank structure on the transition matrices. Many estimation procedures have been proposed including but not limited to $l_1$ regularization \citep{basu2015}, two-stage $l_1$ regularization \citep{davis2016sparse}, sparse seasonal VAR \citep{BAEK2017103}, low rank structured VAR \citep{Basu2018},  hierarchical lag sparsity \citep{nicholson2016, safikhani2018spatio}, banded VAR \citep{band} and nonconcave penalization ( \citealt{zhu2020nonconcave}). Another group of methods assume a factor structure on the time series data to reduce the dimensionality, e.g., \cite{lam2012factor} and \cite{Tu2020factor}. Meanwhile, such high-dimensional techniques become very popular in many applications, such as econometrics (\citealp{MattesonTsay}), genetics \citep{MICHAILIDIS2013}, biology \citep{Hu2019}, ecology \citep{Reyes2012} and so on.

\begin{figure}[h] 
	\begin{minipage}{0.5\linewidth}\centering
		$$\underbrace{\begin{bmatrix} x_{1t}  \\ x_{2t} \\ x_{3t} \\ x_{4t} \\ x_{5t} \end{bmatrix}}_{X_t}
		=
		\underbrace{\begin{bmatrix}  * & * & * & 0 & 0 \\ * & * & 0 & 0 & 0 \\ 0 & 0 & * & 0 & 0 \\
			0 & 0 & * & * & * \\  0 & 0 & 0 & 0 & * \end{bmatrix}}_{\Phi} \underbrace{\begin{bmatrix} x_{1,t-1}  \\ x_{2,t-1} \\ x_{3,t-1} \\ x_{4,t-1} \\ x_{5,t-1} \end{bmatrix}}_{X_{t-1}} + \underbrace{\begin{bmatrix} \epsilon_{1t}  \\ \epsilon_{2t} \\ \epsilon_{3t} \\ \epsilon_{4t}  \\ \epsilon_{5t} \end{bmatrix}}_{\epsilon_t}$$
	\end{minipage}
	\begin{minipage}{0.6\linewidth}\centering
		\includegraphics[width=5.5cm]{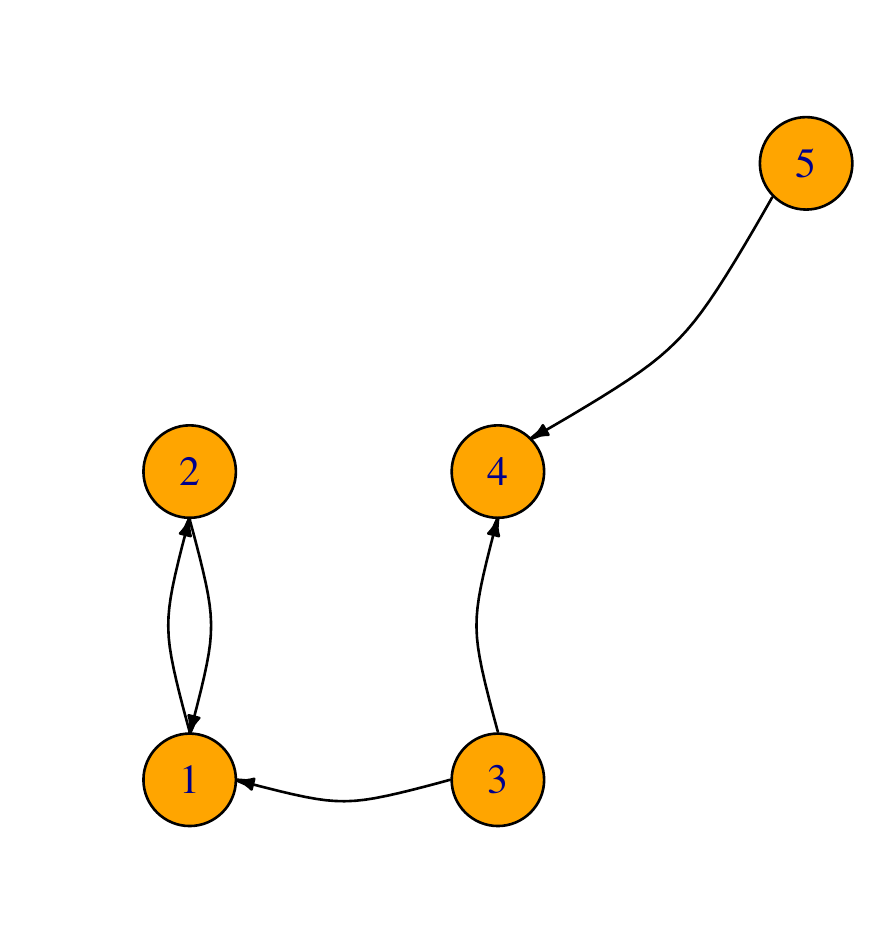}
	\end{minipage}
	\caption{The left panel illustrates the sparsity (zero/non-zero) pattern for the transition matrix $\Phi$ in a VAR(1) process with $*$ denoting non-zero entries. The right panel illustrates the network structure implied by this VAR(1) process. For example, $\Phi_{13}$ is non-zero, which indicates a directed connection from the third site to the first site.}
	\label{fig:netill}
\end{figure}

As for spatio-temporal data, each component of the multivariate time series contains the observations in one spatial location (site). Parameters in the transition matrices can naturally capture the spatial and temporal interrelationship among the sites. Meanwhile the zero-nonzero patterns of the transition matrices reflect the network structure in the dataset. Figure \ref{fig:netill} shows a simple example of VAR(1) model on five sites. We can see there exists a directed connection from site 3 to site 1, indicating that $X_{1t}$ is dependent on $X_{3,t-1}$, so $\Phi_{13}$ is nonzero. Meanwhile $\Phi_{31}=0$ means there is no directed connection from site 1 to site 3. Thus for spatio-temporal data, the spatial structure and temporal information should be incorporated in the modeling procedure. If such information is ignored, high-dimensional methods may lead to inaccurate network estimation and unreasonable scientific conclusion. Figure \ref{fig:netill} illustrates the drawback of ignoring the spatial and temporal information based on a simulation study in Section \ref{subsec:simuVAR1}, in which the blue edges and red edges stand for false negatives and false positives respectively. Without considering the spatial and temporal information,  LASSO not only underestimates true connections but also overestimates wrong connections, while nonconcave penalized estimation (SCAD and MCP) severely underestimate the true connections. In contrast, our proposed method (WLASSO1 and WLASSO2) recovers the network very well and significantly reduces false positives and false negatives.

\begin{figure}[!h]
	\centering
		\includegraphics[width=0.8\textwidth]{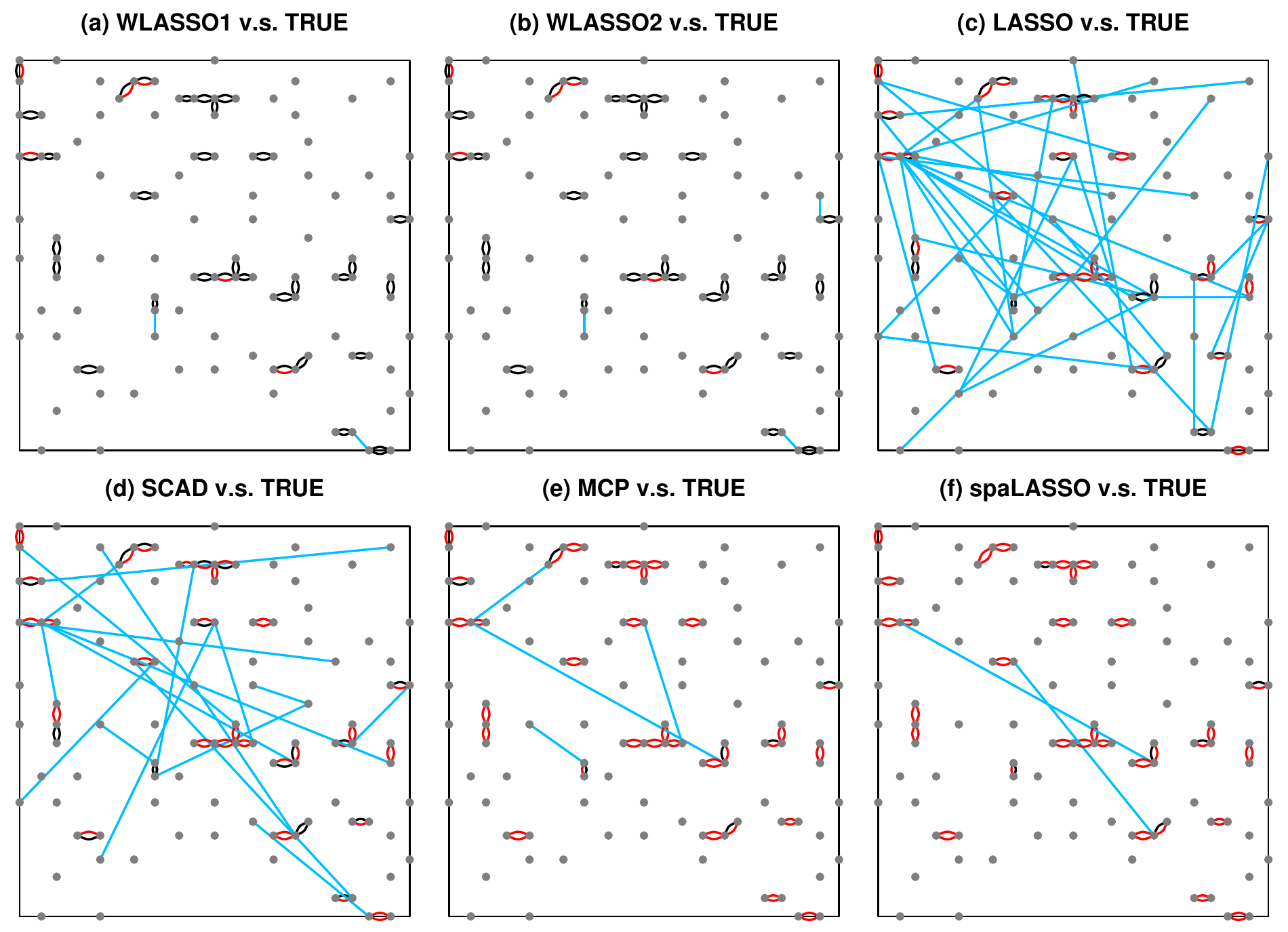} 
	\caption{Comparison of the proposed methods (WLASSO1 and WLASSO2) to four existing methods (LASSO, SCAD, MCP and spaLASSO from \cite{schweinberger2017high}) in network estimation of one simulated VAR(1) process from Section \ref{subsec:simuVAR1}. In detail, if both true value $\Phi_{ss'}$ and its estimator $\hat{\Phi}_{ss'}$ are nonzero, a black edge is drawn to connect site $s$ and site $s'$. If $\Phi_{ss'}$ is not zero but $\hat{\Phi}_{ss'}$ is zero, the edge is red. If $\Phi_{ss'}$ is zero but $\hat{\Phi}_{ss'}$ is not zero, the edge is blue.}
	\label{fig:net_ill}
\end{figure}

In this paper, we proposed a data-driven weighted $l_1$ regularized approach that constructs the penalty according to spatial distance among sites and temporal lags in the VAR model. We derived non-asymptotic upper bounds of the estimation error which hold with high probability, and showed these bounds are smaller than those from LASSO (remark (c) in Section \ref{subsec:exact} and Section \ref{subsec:weak}). The simulation studies compare the proposed approach with four existing methods for high-dimensional VAR including LASSO \citep{basu2015}, SCAD and MCP \citep{zhu2020nonconcave}, and spaLASSO \citep{schweinberger2017high}. The proposed approach shows significant advantage in model fitting, 
network detection
and forecasting performance (Table \ref{table:simu1}-\ref{table:simu_fitp3} and Figure \ref{fig:net_p2}-\ref{fig:forecast_p23} in the supplemental material). 
We applied our method to a traffic network dataset in Des Moines, Iowa area. The network structure detected by LASSO was not meaningful while the proposed method provides much more reasonable estimated network and better forecasting results. 

There are few papers focusing on high-dimensional VAR in the spatio-temporal setting. The most relevant one is \cite{schweinberger2017high}, denoted as spaLASSO, which incorporates spatial structure in VAR model estimation. Their approach assumes the spatial dependence only exists within a specific distance $\rho$, while $\rho$ is either known or estimated in an initial step by LASSO within sub-sampled sites. After $\rho$ is specified, only  parameters associated with distances smaller than the given $\rho$ are estimated, and others are fixed as zero. Assuming the distance $\rho$ is known is usually unrealistic in real data sets. In estimating $\rho$ by an initial LASSO estimator, inaccuracy of the initial estimator can produce unreliable estimation of $\rho$, thus contaminating the final estimation of the  model.  As shown in Figure \ref{fig:net_ill} (c), LASSO cannot identify the true network and therefore would deliver inaccurate estimation of $\rho$ and eventually results in an inaccurate estimation from spaLASSO (Figure \ref{fig:net_ill} (f)). Further, the assumption of no spatial dependence beyond distance $\rho$ is restrictive, and may not be true in some real cases, such as the more general weakly sparse scenario considered in this paper. In addition, this approach also does not incorporate the lag order of temporal dependence. 
	In contrast, the proposed method incorporates both spatial and temporal information in a smooth way rather than truncating the parameters at a certain distance, and the penalty weights are data-driven so that no prior information is needed. The algorithm of the proposed method is one-step and easy to carry out via existing algorithms.
	
	In real application, spatial and temporal dependence may still exist even for a long distance or temporal lag. In such cases, the transition matrices in the VAR model will have many small non-zero elements thus are not sparse, which is the so called ``weakly sparse" scenario. The second goal of this paper is to investigate the theoretical properties of $l_1$ regularized estimation of VAR model under weakly sparse scenario. Weak sparsity is pursued mostly for independent data including \cite{negahban2009unified} and \cite{Raskutti2011}. There is a gap in the literature in investigating the properties of $l_1$ regularized estimation for high-dimensional VAR models under the weakly sparse scenario. Our contribution is to fill this gap. In addition, the ``weak sparsity" defined in this paper is more general than the $l_r$ ball constraint which is commonly used in literature and we will discuss the advantages of our weak sparsity in detail in Section. 
	We first derived the upper bounds of $l_1$ regularized estimation error for general stationary VAR process (Theorem \ref{weaksp_thm}) and provided the weak sparsity constraint (\ref{weaksp_def}) which guarantees the estimation consistency. Then we further explored the weak sparsity constraint and simplified it in a special case of VAR(1) process. Moreover, the results in Theorem \ref{weaksp_thm} can also be directly used to derive the error bound under the $l_r$ ball setting (Corollary \ref{weaksp_coro}) 
	and we proved our weak sparsity constraint is more relaxed than the $l_r$ ball setting (Remark (a) of Corollary \ref{weaksp_coro}). Finally, the proposed method under the weakly sparse scenario is examined in the simulation studies, which shows impressive advantages over other existing methods.

	\paragraph{Outline of the Paper:} The remainder of the paper is structured as follows. Section \ref{sec:esti} introduces the weighted $l_1$ regularized approach for high-dimensional spatio-temporal VAR and its theoretical properties. Section \ref{sec:simu} presents the implementation of the proposed method and compares its performance with LASSO, SCAD, MCP and spaLASSO through several simulation studies. Application on the traffic network dataset is in Section \ref{sec:realdat}, followed by the conclusion in Section \ref{sec:conclusion}.

\paragraph{Notation:} Throughout this paper, we denote the cardinality of a set $J$ by $|J|$, and use $J^C$ to denote its complementary set. For a vector $\bv\in R^n$, we use $\bv_J:=(\bv_i)_{i\in J}$ to denote the sub-vector with support $J$, and use $\norm{\bv}_q\defeq (\sum_{i=1}^n|v_i|^q)^{1/q}$ to denote its $l^q$ norm. For a matrix $A$, we use $A_j$ to denote its $j$th column, $\tv(A)$ to denote its vectorization, $A'$ and $A^H$ are its transpose and conjugate transpose respectively. $A\circ B$ and $A\otimes B$ are the element-wise product and Kronecker product of matrices $A$ and $B$ respectively.  For a symmetric or Hermitian matrix $A$, $\Lambda_{\max}(A)$ and $\Lambda_{\min}(A)$ are the largest and smallest eigenvalue respectively. For a squared matrix $A$, we use $\norm{A}_F$, $\rho(A)$ and $\norm{A}_2$ to denote its Frobenius norm $\sqrt{tr(A^HA)}$, spectral radius $\max\{|\lambda_i|: \lambda_i \text{'s are eigenvalues of } A\}$, and spectral norm $\sqrt{\Lambda_{\max}(A^HA)}$ respectively. For convenience, we use $\bone_{n}$ and $\bzero_n$ to denote a vector of 1 and 0 with length $n$ respectively. We write $x \gtrsim y$ if there exists a positive constant $c$, such that $x\geq cy$. If we have both $x \gtrsim y$ and $y \gtrsim x$, we use $x\asymp y$ to denote their relationship.

\section{High-Dimensional Spatio-Temporal Vector Autoregression}
\label{sec:esti}
Suppose $x_{st}$ is the observation on site $s$ at time $t$ ($s=1,\cdots,m$; $ t=1,\cdots,T$), and we assume $X_t=(x_{1t},\cdots,x_{mt})'$ is generated by a $p$-th order vector auto-regressive (VAR) process:
\begin{equation}
X_t=\Phi_1X_{t-1}+\cdots+\Phi_pX_{t-p}+\varepsilon_t, \quad \varepsilon_t\overset{\tt{i.i.d}}{\sim} N(\bzero, \Sigma),
\label{eq:var}
\end{equation}
Here $\Phi_1,\cdots,\Phi_p$ are $m\times m$ transition matrices encoding dependence across space and temporal lags. We use $\Phi_{l,ss'}$ to denote the $ss'$-th entry of $\Phi_l$, so that $\Phi_{l,ss'}$ represents the $l$-lagged influence of site $s'$ on site $s$. We express this VAR($p$) model as the following multivariate regression form:
\vskip -12pt
\begin{eqnarray*}
	\underbrace{\begin{bmatrix}
			X'_T \\ \vdots \\ X'_{p+1}
	\end{bmatrix}}_{\bY_{(T-p)\times m}}
	&=& 
	\underbrace{\begin{bmatrix}
			X'_T & \cdots & X'_{T-p}\\ 
			\vdots & \ddots & \vdots \\
			X'_p & \cdots & X'_1
	\end{bmatrix}}_{\bX_{(T-p)\times pm}}
	\underbrace{\begin{bmatrix}
			\Phi'_1 \\ \vdots \\ \Phi'_p 
	\end{bmatrix}}_{\bB_{pm\times m}}
	+
	\underbrace{\begin{bmatrix}
			\varepsilon_T \\ \vdots \\ \varepsilon_{p+1} 
	\end{bmatrix}}_{\bE_{(T-p)\times m}}.
	\\                
\end{eqnarray*}
\vskip -8pt
In the high-dimensional case, LASSO can recover the sparseness of transition matrices and reduce forecasting error \citep{basu2015}.
However, regular LASSO uses the same penalty for different $\Phi_{l,ss'}$ components, which may be  inappropriate for spatio-temporal data. Instead, we proposed the following weighted $l_1$ regularized LS, which penalizes $\Phi_{l,ss'}$ differently according to the spatial distance between site $s$ and $s'$, say $d_{ss'}$, as well as the temporal lag $l$:
\begin{equation}
\text{weighted $l_1$-LS:} \quad \widehat{\bB} = \min_{\bB}\; \frac{1}{N}\norm{\bY-\bX \bB}_F^2 +\lambda_N\Omega(\bB),
\label{eq:wls} 
\end{equation}
where $N=T-p$ and $\Omega(\bB)=\sum_{l=1}^p\sum_{s,s'=1}^m w_{l,ss'}|\Phi_{l,ss'}|$ with $ w_{l,ss'}\geq 0$ being the penalty weight for $\Phi_{l,ss'}$. Since $\Phi_{l,ss'}$ quantifies the dependence between site $s$ and site $s'$ across temporal lag $l$, it is more likely to be zero if $d_{ss'}$ and $l$ are large. Therefore, the weight $ w_{l,ss'}$ is set to be an increasing function of distance $d_{ss'}$ and temporal lag $l$. Through this construction of penalty weights we impose a spatio-temporal structure on the data in that the conditional dependence among two sites across temporal lag $l$ (represented by $\Phi_{l,ss'}$) decays as spatial distance $d_{ss'}$ and temporal lag $l$ increase. There are several ways to define the weights, for example:
\begin{equation}
	w_{l,ss'}^{(1)}= \exp \left( c_1 \frac{l \, d_{ss'}}{p\,d_{max}} \right)\quad \text{or} \quad w_{l,ss'}^{(2)} = \left(1+ \frac{l \, d_{ss'}}{p\,d_{max}}\right)^{c_2},
	\label{eq:weight}
\end{equation}
where $d_{max}$ is the maximum of $d_{ss'}$  
and $c_1,c_2>0$ are universal constants to be determined by cross validation. The inclusion of $c_1$ and $c_2$ ensures weights are data-driven and adds flexibility to this method. Other weight functions can be defined as well based on the context of the dataset under investigation. A special case is that $w_{l,ss'}$ is only a function of $d_{ss'}$ such as $w_{l,ss'}^{(3)}= \exp \left( c_3 d_{ss'}/d_{max} \right)$, which means the magnitudes of parameters are only influenced by the distance. The performances of different weight functions are examined in simulation studies and real data application. 

Utilizing weighted penalty functions such as those above significantly improves model performance without sensitivity to the exact choice of weight functions. This is mainly due to including a data-informed constants $c_i$ in all weight functions, which are selected via cross-validation. Including such data-driven constants optimizes the weight to some extent and reduces the reliance of model performance on the choice of weight function, demonstrating the robustness of the proposed method with respect to changes in the weight functions.


\subsection{Model Assumption}
\label{sec:theory}
In the following, we provide non-asymptotic bounds on the estimation error of weighted $l_1$-LS estimation (\ref{eq:wls}), and show that under certain conditions the proposed estimator is consistent. We rewrite the VAR model as:
$$\tv(\bY) = \tv(\bX\bB) +  \tv(\bE) = (I_m\otimes \bX) \tv(\bB) + \tv(\bE) := \bZ \bbeta + \be, $$
where $\by=\tv(\bY)$ is $mN\times 1$ vector, $\bZ=I_m\otimes \bX$ is $mN\times q$ matrix and $\bbeta=\tv(\bB)$ is $q\times 1$ vector with $q=m^2p$. The proposed estimation (\ref{eq:wls}) can be expressed as the following M-estimation:
\begin{equation}
\hbeta = \argmin_{\bbeta}\; \lbk -2\bbeta'\hgamma + \bbeta'\hGamma\bbeta +\lambda_N\Omega(\bbeta)\rbk,
\label{eq:Mest}
\end{equation}
where $\hgamma=(I_m\otimes\bX')\by/N$ and $\hGamma=(I_m\otimes\bX'\bX)/N$. Throughout this paper, we denote the true parameter as $\bbeta^*$ and the corresponding true transition matrices as $\Phi^*_1,\cdots,\Phi^*_p$. We consider two scenarios: (1) $\bbeta^*$ is exactly sparse; (2) $\bbeta^*$ is not exactly sparse, but can be well approximated by a sparse vector, which is called ``weakly sparse". Both scenarios need the following assumption:

\begin{assumption}
	VAR($p$) process is stationary, that is, the roots of $|I_m-\sum_{l=1}^p\Phi_l z|=0$ are lying outside the unit circle. Also $\Sigma$ is positive definite. 
	\label{stable_ass}
\end{assumption} 

This is a fundamental assumption in high-dimensional time series analysis. Since the key in analyzing the M-estimation (\ref{eq:Mest}) is the dependence shown in $\hgamma$ and $\hGamma$, this assumption guarantees that the spectral density of $\{X_t\}$ exists. Under such assumption, \cite{basu2015} used spectral density to construct measure of dependence and proved that $\hgamma$ and $\hGamma$ satisfy two important conditions. More specifically, Proposition (4.2) and (4.3) in \cite{basu2015} state that, under Assumption \ref{stable_ass}, there exist constants $b_i$, such that for $N \gtrsim \max\{\omega^2,1\}(\log p+2\log m)$, the RE condition (\ref{eq:RE}) and Derivation condition (\ref{eq:der}) hold with probability at least $1-b_1\exp(-b_2 N\min\{\omega^{-2},1\})-b_3\exp(-b_4(\log p + 2\log m))$:
\begin{align}
\text{Restricted Eigenvalue (RE):} \quad &\theta'\hGamma\theta\geq \alpha\norm{\theta}_2^2 - \tau\norm{\theta}_1^2, \quad \forall \theta\in R^q,       \label{eq:RE} \\
\text{Derivation condition:}\quad & \|\hgamma-\hGamma\bbeta^*\|_{\infty}\leq\bQ\sqrt{\frac{\log p + 2\log m}{N}}. \label{eq:der}
\end{align}
Here $\omega$, $\alpha$, $\tau$ and $\bQ$ are determined by the transition matrices $\{\Phi^*_l\}_{l=1}^p$ and covariance matrix of the innovation $\Sigma$. In details, first we define $$\mu_{\min}(\bPhi)=\underset{|z|=1}{\min}\;\Lambda_{\min}(\bPhi^H(z)\bPhi(z)), \quad \mu_{\max}(\bPhi)=\underset{|z|=1}{\max}\;\Lambda_{\max}(\bPhi^H(z)\bPhi(z)),$$ 
	where $\bPhi(z)=I-\sum_{l=1}^p\Phi^*_lz^l$ ($z\in\mathbb{C}$) is the characteristic polynomial of the VAR process and $\bPhi^H(z)$ is its conjugate transpose. Further we set 
\begin{align*}
  \tilde{\Phi}=\begin{bmatrix}\Phi_{1} & \cdots & \Phi_{p-1} & \Phi_{p}\\
		I_{m} & \cdots & \bzero & \bzero \\
		\vdots & \ddots & \vdots & \vdots\\
		\bzero & \cdots & I_{m} & \bzero
	\end{bmatrix},\quad
  \begin{array}{l@{\mskip\thickmuskip}l}
    \tilde{\bPhi}(z)=I_{pm}-\tilde{\Phi}(z) \;(z\in\mathbb{C}), \\
    \mu_{\min}(\tilde{\bPhi})=\underset{|z|=1}{\min}\;\Lambda_{\min}(\tilde{\bPhi}^H(z)\tilde{\bPhi}(z)).
  \end{array}
\end{align*}
	Then $\omega$, $\alpha$, $\tau$ and $\bQ$  are defined as follows:
	$$\omega=a_1\frac{\Lambda_{\max}(\Sigma)/\mu_{\min}(\tilde{\bPhi})}{\Lambda_{\min}(\Sigma)/\mu_{\max}(\bPhi)}, \quad \alpha=\frac{\Lambda_{\min}(\Sigma)}{2\mu_{\max}(\bPhi)},\quad \tau=\alpha\max\{\omega^2,1\}\frac{\log p+\log m}{N},$$ $$\bQ=a_2\lmk \Lambda_{\max}(\Sigma) + \frac{\Lambda_{\max}(\Sigma)}{\mu_{\min}(\bPhi)} + \frac{\Lambda_{\max}(\Sigma)\mu_{\max}(\bPhi)}{\mu_{\min}(\bPhi)} \rmk,$$ 
	where $a_1$ and $a_2$ are positive constants. We refer to \cite{basu2015} for more details. The RE condition (\ref{eq:RE}) and Derivation condition (\ref{eq:der}) are the key to derive convergence rate of the M-estimation (\ref{eq:Mest}). 

\subsection{Convergence Rate under Exact Sparsity}
\label{subsec:exact}
In this section, we assume the true parameter $\bbeta^*$ has many zero entries, and we set its support to be $J=\{(l,ss'): \Phi^*_{l,ss'}\neq 0\}$ with $|J|=k$. Further we need the following constraint for the penalty weights:

\begin{assumption}
	$w_{l,ss'} >0$ for all $(l,ss')\in J^C$.
	\label{exactsp_ass}
\end{assumption} 

This assumption states that the parameters with true values being zero should have nonzero penalties. This assumption can be guaranteed by setting all penalty weights to be positive. 
In addition, any choice of $(\lambda_N,\{w_{l,ss'}\})$ is equivalent to $(\tilde{\lambda}_N,\{\tilde{w}_{l,ss'}\})$ with $\tilde{\lambda}_N=a\lambda_N$ and $\tilde{w}_{l,ss'}=w_{l,ss'}/a$ for any arbitrary positive number $a$. So without lost of generality, we can set $\min\{w_{l,ss'}:(l,ss')\in J^C\}=1$. Further we set $r_w=\max\{w_{l,ss'}:(l,ss')\in J\}$, which is indeed the ratio between the maximum weight of nonzero parameters and the minimum weight of zero parameters, i.e. $r_w=\max\{w_{l,ss'}:(l,ss')\in J\}/\min\{w_{l,ss'}:(l,ss')\in J^C\}$. In the following theorem, we can see this ratio is the key quantity for the proposed method to achieve smaller error bounds than LASSO.

\begin{theorem}
	Consider weighted $l_1$-LS estimator in (\ref{eq:Mest}). If Assumption \ref{stable_ass} and \ref{exactsp_ass} hold, there exist constants $b_i>0$ not depending on data and model parameters, such that for any $N \gtrsim (1+r_w)^2\max\{\omega^2,1\}k(\log p+2\log m)$ and $\lambda_N\geq 4\bQ\sqrt{(\log p + 2\log m)/N}$, with at least probability:
	$$1-b_1\exp(-b_2N\min\{\omega^{-2},1\})-b_3\exp(-b_4(\log p + 2\log m)),$$
	the estimation error $(\hbeta-\bbeta^*)$ is bounded as follows:
	\begin{eqnarray*}
	&&\|\hbeta-\bbeta^*\|_2 \leq \frac{1+2r_w}{\alpha}\sqrt{k}\lambda_N, \quad \|\hbeta-\bbeta^*	\|_1 \leq \frac{2+6r_w+4r_w^2}{\alpha}k\lambda_N, \\
	&&(\hbeta-\bbeta^*)'\hGamma (\hbeta-\bbeta^*) \leq \frac{(1+2r_w)^2}{2\alpha}k\lambda_N^2.
	\end{eqnarray*}
	If we set $s_0=\min\{|\beta^*_j|: j\in J\}$, the number of false zero is bounded by:
	\begin{equation*}
	\left| \mathrm{supp}(\bbeta^*)\backslash \mathrm{supp}(\hbeta)\right|  \leq \frac{2+6r_w+4r_w^2}{s_0\alpha}k\lambda_N.
	\end{equation*}
	If we consider a threshold version $\tilde{\bbeta}:=\{\hat{\beta}_{j}I(|\hat{\beta}_j|>\lambda_N)\}$ with $I(.)$ being the indicator function, the number of false non-zero in $\tilde{\bbeta}$ is bounded by: 
	\begin{equation*}
	\left| \mathrm{supp}(\tilde{\bbeta})\backslash \mathrm{supp}(\bbeta^*)\right|  \leq (1+2r_w)^2\frac{k}{\alpha}.
	\end{equation*}
	\label{exactsp_thm}
\end{theorem}
\vspace{-1.3cm}
\paragraph*{Remarks.} 
(a) $\|\hbeta-\bbeta^*\|_2=\sqrt{\sum_{l=1}^p\|\hat{\Phi}_l-\Phi^*\|^2_F}$ is the error of transition matrices under Frobenius norm. 
$(\hbeta-\bbeta^*)'\hGamma (\hbeta-\bbeta^*)=\sum_{t=1}^T\| \sum_{l=1}^p (\hat{\Phi}_l-\Phi^*)X_{t-l}\|_2^2/T$ is the in-sample prediction error under $l_2$ norm. 

(b) If we set $r_w=1$ which corresponds to LASSO, we will get the following upper bounds that are similar to those in \cite{basu2015}: $\|\hbeta-\bbeta^*\|_2 \leq 3\sqrt{k}\lambda_N/\alpha$, $\|\hbeta-\bbeta^*\|_1 \leq 12k\lambda_N/\alpha$, $(\hbeta-\bbeta^*)'\hGamma (\hbeta-\bbeta^*) \leq 9k\lambda_N^2/\alpha$,
$|\mathrm{supp}(\bbeta^*)\backslash \mathrm{supp}(\hbeta)  \leq 12k\lambda_N/(s_0\alpha)$, $\left| \mathrm{supp}(\tilde{\bbeta})\backslash \mathrm{supp}(\bbeta^*)\right|  \leq 9k/\alpha$.

(c) Compared with LASSO ($r_w=1$),  if weights $\{w_{l,ss'}\}$ are properly specified, the ratio $r_w$ should be much smaller than one. In the ideal case when $r_w$ is close to zero, our upper bounds for $\|\hbeta-\bbeta^*\|_2$, $\|\hbeta-\bbeta^*\|_1$, $(\hbeta-\bbeta^*)'\hGamma (\hbeta-\bbeta^*)$, $\left| \mathrm{supp}(\bbeta^*)\backslash \mathrm{supp}(\hbeta)\right|$ and $\left| \mathrm{supp}(\tilde{\bbeta})\backslash \mathrm{supp}(\bbeta^*)\right|$ 
are nearly 1/3, 1/6, 1/9, 1/6 and 1/9 of that from LASSO respectively. 

(d) Condition of Consistency: Since the upper bound of $l_2$ error holds with probability converging to one, $\sqrt{k}\lambda_N/\alpha\rightarrow 0$ is sufficient to gain consistency of $\hbeta$. Furthermore, if we set $\lambda_N\asymp \bQ\sqrt{(\log p + 2\log m)/N}$, in special cases when $\bQ/\alpha$ is bounded away from infinity, we have $\|\hbeta-\bbeta^*\|_2 \lesssim \sqrt{k(\log p + 2\log m)/N}$. Thus the consistency only requires that $N$ increases at a faster rate than $k(\log p+2\log m)$.

\subsection{Convergence Rate under Weak Sparsity}
\label{subsec:weak}
In real applications, the conditional dependence quantified by $\Phi^*_{l,ss'}$ may not be zero even for large distance $d_{ss'}$ and/or lag $l$. For example, if the underlying true process is a Vector Auto-regressive Moving Average (VARMA) process but we use VAR to approximate it, $\Phi^*_{l,ss'}$ is generally nonzero for large $l$. Also, $\Phi^*_{l,ss'}\neq 0$ may occur for large distance $d_{ss'}$ especially when the sites are located on an irregular lattice. These examples motivate us to consider a scenario called ``weak sparsity", in which the true parameter vector $\bbeta^*$ does not have many zeros (i.e. not exactly sparse) but can be well approximated by a sparse vector. There are only few results in the literature discussing weak sparsity, and almost all of them have focused on independent data (\citealp{negahban2009unified}, \citealp{Raskutti2011}), except \cite{Basu2018spectral} which focuses on estimating the spectral density matrix of high-dimensional time series. Moreover, they each define weak sparsity under the so-called `` $l_r$ ball" setting. Specifically, they assumed the true parameter vector is within the $l_r$ ball: $\mathbb{B}_r(R):=\lbk \bbeta^*: \sum_{j=1}^q|\beta_j^*|^r \leq R \rbk$ where $r\in [0,1]$ is fixed. In this setting, a constraint on the radius $R$ is required to achieve the estimation consistency. For example, in independent data, LASSO estimator is consistent if $R$ satisfies:   
\begin{equation}
  l_r \text{ ball constriant:}\quad R = o\lsk \lsk\frac{N}{\log q}\rsk^{1-r/2} \rsk,
  \label{eq:lrball_indep}
\end{equation}
where $q$ is the number of parameters (\citealp{negahban2009unified}, \citealp{Raskutti2011}). However, how ``sparsifiable" $\bbeta^*$ is depends on the relative magnitude of each element in $\bbeta^*$ rather than its overall $l_r$ length. Thus the $l_r$ ball setting does not clearly describe the ``sparsifiablility" of $\bbeta^*$. A special case in which all $\beta_j^*$s have the same magnitude could still fit in the $l_r$ ball setting. While, in this case $\bbeta^*$ cannot be approximated by a sparse vector and is not suitable for $l_1$ regularized estimation.  As a consequence, in general the $l_r$ ball setting may not be a reasonable way to relax the sparsity assumption.


Instead of using the $l_r$ ball setting, we define ``weak sparsity'' from another perspective: most entries of $\bbeta^*$ are small enough such that $\bbeta^*$ can be well approximated by its hard thresholding version, say $\bbeta^*_{\eta}$, whose $j$th entry is $\beta^*_jI(|\beta^*_j|>\eta)$. For any given threshold $\eta$, we use $J_{\eta}=\{j: |\beta^*_j|>\eta\}$ to denote the support of $\bbeta^*_{\eta}$. The formal definition of our proposed weak sparsity is as follows. \\
{\bf Definition (Weak Sparsity Constraint):} If there exists an $\eta$ such that the following two conditions hold,
\vspace{-12pt}
\begin{align}
&|J_{\eta}|=o\lsk\lsk\frac{\alpha}{\bQ}\rsk^2\frac{N}{\log p+2\log m}\rsk\nonumber\quad \text{and} \\ &\|\bbetaJC\|_1=o\lsk\min\lbk\frac{\alpha}{\bQ},1,\frac{1}{\omega}\rbk\sqrt{\frac{N}{\log p+2\log m}}\rsk
\label{weaksp_def}
\end{align}
where $J_{\eta}^C:=\{j: |\beta_j^*|\leq \eta\}$, we say $\bbeta^*$ satisfies the weak sparsity constraint.

This constraint means, with a proper choice of $\eta$, $\bbeta^*_{\eta}$ is sparse and is a good approximation of $\bbeta^*$ in the sense that its difference from $\bbeta^*$, denoted as $\bbetaJC$, is small enough. In this way, our weak sparsity constraint quantifies how sparsifiable the true parameter vector $\bbeta^*$ is so that its $l_1$ regularized estimation remains consistent. In the following theorem, first without this constraint, we give a general result of the upper bound of the estimation error. Then under this weak sparsity constraint, with proper choice of $\lambda_N$ we can show the proposed estimator is consistent. Furthermore, we simplified the weak sparsity constraint in a special case of VAR(1) in Proposition \ref{weaksp_prop}. Finally, we directly apply Theorem \ref{weaksp_thm} to derive the upper bound of estimation error under the $l_r$ ball setting and prove our weak sparsity constraint (\ref{weaksp_def}) is more relaxed than the $l_r$ ball constraint(Corollary \ref{weaksp_coro}).  Also notice that the following Theorem \ref{weaksp_thm} and Corollary \ref{weaksp_coro} also hold for LASSO, since LASSO can be viewed as a special case of the proposed method where all $w_{l,ss'}$s are the same. To state our theorem, we define the following notations: for any $\eta$, we set $w_1(\eta)=\min\{w_{l,ss'}: (l,ss')\in J^C_{\eta}\}$, $w_2(\eta)=\max\{w_{l,ss'}: (l,ss')\in J_{\eta}\}$ and $r_w(\eta)=w_2(\eta)/w_1(\eta)$.

\begin{assumption}
	$w_{l,ss'}>0$ for all $(l,ss')$.
	\label{weaksp_ass}
\end{assumption}  

\begin{theorem}
	Consider weighted $l_1$-LS estimator in (\ref{eq:Mest}) and assume Assumption \ref{stable_ass} and \ref{weaksp_ass} hold. Then there exist constants $b_i>0$, such that for any $\eta$, if $N \gtrsim (1+r_w(\eta))^2|J_{\eta}| \max\{\omega^2,1\}(\log p+2\log m)$ and $\lambda_N= \widetilde{\lambda}_N/w_1(\eta)$ with $\widetilde{\lambda}_N=4\bQ\sqrt{(\log p + 2\log m)/N}$, with at least probability:
	$$1-b_1\exp(-b_2N\min\{\omega^{-2},1\})-b_3\exp(-b_4(\log p + 2\log m)),$$
	the estimation error $(\hbeta-\bbeta^*)$ will be bounded as follows:
	\begin{eqnarray*}
	\|\hbeta-\bbeta^*\|_2 &\leq &  \frac{1+2r_w(\eta)}{\alpha}\sqrt{|J_{\eta}|}\widetilde{\lambda}_N + 2\sqrt{\frac{r_w(\eta)\widetilde{\lambda}_N\|\bbeta^*_{J^C_\eta}\|_1}{\alpha}} + \\
	&&\frac{4r_w(\eta)\max\{\omega,1\}}{\bQ} \widetilde{\lambda}_N \|\bbeta^*_{J^C_\eta}\|_1 , \\
	\|\hbeta-\bbeta^*\|_1 &\leq & (2+r_w(\eta))\sqrt{|J_{\eta}|}\|\hbeta-\bbeta^*\|_2 + 4r_w(\eta)\|\bbetaJC\|_1, \\
	(\hbeta-\bbeta^*)'\hGamma (\hbeta-\bbeta^*) &\leq & \frac{1+2r_w(\eta)}{2}\sqrt{|J_{\eta}|}\widetilde{\lambda}_N\|\hbeta-\bbeta^*\|_2 + 2r_w(\eta)\widetilde{\lambda}_N\|\bbetaJC\|_1.
	\end{eqnarray*}
	Secondly, if there exists an $\eta$ such that $\bbeta^*$ satisfies the weak sparsity constraint (\ref{weaksp_def}),  the proposed estimator is consistent, i.e. for any arbitrary $\epsilon>0$, $Pr\lsk\|\hbeta-\bbeta^*\|_2 >\epsilon\rsk\rightarrow 0$ as $T,m\rightarrow\infty$.\\
	\label{weaksp_thm}
\end{theorem}
\vspace{-1.3cm}
\paragraph*{Remarks:} (a) Theorem \ref{weaksp_thm} includes the exact sparsity as a special case. If $\bbeta^*$ is exactly sparse with $k$ nonzero entries, by setting $\eta=0$ we can obtain $|J_{\eta}|=k$ and $\|\bbetaJC\|_1=0$. Then the above three upper bounds are exactly the same as those in Theorem \ref{exactsp_thm}. For weakly sparse scenario, we approximate $\bbeta^*$ by its hard thresholding version $\bbeta^*_{\eta}$. As a consequence, extra terms containing $\|\bbetaJC\|_1$ occur in the upper bounds. 

(b) By setting $r_w=1$, we can obtain the upper bounds of LASSO:
	\begin{eqnarray*}
	&& \|\hbeta-\bbeta^*\|_2 \leq  \frac{3}{\alpha}\sqrt{|J_{\eta}|}\widetilde{\lambda}_N + 2\sqrt{\frac{\widetilde{\lambda}_N\|\bbeta^*_{J^C_\eta}\|_1}{\alpha}} + \frac{4\max\{\omega,1\}}{\bQ} \widetilde{\lambda}_N \|\bbeta^*_{J^C_\eta}\|_1 , \\
	&&\|\hbeta-\bbeta^*\|_1 \leq 3\sqrt{|J_{\eta}|}\|\hbeta-\bbeta^*\|_2 + 4\|\bbetaJC\|_1, \\
	&& (\hbeta-\bbeta^*)'\hGamma (\hbeta-\bbeta^*) \leq \frac{3}{2}\sqrt{|J_{\eta}|}\widetilde{\lambda}_N\|\hbeta-\bbeta^*\|_2 + 2\widetilde{\lambda}_N\|\bbetaJC\|_1.
\end{eqnarray*}
Further, if the weak sparsity constraint (\ref{weaksp_def}) holds, LASSO estimator is also consistent.  

(c) If weights $\{w_{l,ss'}\}$ are properly specified, ratio $r_w$ should be smaller than one and implies smaller error bounds comparing with LASSO. In the ideal case when $r_w$ is close to zero, the error bounds of the proposed method are approaching to:
	\begin{eqnarray*}
		&& \|\hbeta-\bbeta^*\|_2 \leq \frac{1}{\alpha}\sqrt{|J_{\eta}|}\widetilde{\lambda}_N, \quad \|\hbeta-\bbeta^*\|_1 \leq 2\sqrt{|J_{\eta}|}\|\hbeta-\bbeta^*\|_2, \\
		&& (\hbeta-\bbeta^*)'\hGamma (\hbeta-\bbeta^*) \leq \frac{1}{2}\sqrt{|J_{\eta}|}\widetilde{\lambda}_N\|\hbeta-\bbeta^*\|_2,
	\end{eqnarray*}
	which are less than 1/3, 2/9 and 1/9 of those from LASSO respectively. 

The meaning of the weak sparsity constraint (\ref{weaksp_def}) is straightforward. However, it is  hard to verify in application since it contains $\alpha$, $\bQ$ and $\omega$ which depend on unknown model parameters. When $\alpha$ is bounded away from zero, $\bQ$ and $\omega$ are bounded away from infinity, this constraint can be simplified as $|J_{\eta}|=o\lsk N/(\log p+2\log m)\rsk$ and $\|\bbetaJC\|_1=o\lsk N/(\log p+2\log m)\rsk$, which only depends on the number of observation and parameter dimension. For general stationary VAR process, the behaviors of $\alpha$, $\bQ$ and $\omega$ are complex and cannot be guaranteed to be bounded. Here we consider a simple case of VAR(1) process whose transition matrix is symmetric, and explore the properties of $\alpha$, $\bQ$ and $\omega$ in the following Proposition \ref{weaksp_prop}. 

\begin{proposition}
	For any stationary VAR(1) process $X_t=\Phi X_{t-1}+\epsilon_t$ whose transition matrix $\Phi$ is symmetric, we have
	\begin{eqnarray*}
		&&|\lambda_i|<1 \; \text{for any } i,\quad \rho(\Phi)=\underset{1\leq i\leq m}{\max} |\lambda_i|, \label{eq:eigen}\\
	&&\mu_{\max}(\bPhi)=(1+\rho(\Phi))^2,\; \mu_{\min}(\bPhi)=(1-\rho(\Phi))^2,
	\label{eq:mu}
	\end{eqnarray*}
	where $\{\lambda_i\}_{i=1}^m$ are the eigenvalues of $\Phi$. Furthermore, $\alpha$ is bounded away from zero, $\bQ$ and $\omega$ are bounded away from infinity if and only if $\Lambda_{\max}(\Sigma)$ is bounded away from infinity, $\Lambda_{\min}(\Sigma)$ is bounded away from zero and $\rho(\Phi)$ is bounded away from 1. 
	\label{weaksp_prop}
\end{proposition}
This proposition implies: for VAR(1) process with symmetric transition matrix, if the eigenvalues of $\Sigma$ and $\Phi$ behave properly and there exists an $\eta>0$ satisfying weak sparsity constraint (\ref{weaksp_def}), we will achieve the consistency of $\hbeta$.
\paragraph{$l_r$ Ball Setting:} \cite{negahban2009unified} and \cite{Raskutti2011} investigate LASSO estimation of linear regression in independent data under the $l_r$ ball setting. Under some conditions, they built up the upper bound of $l_2$ estimation error and provided the condition of consistency (i.e. $l_r$ ball constraint (\ref{eq:lrball_indep})). Based on Theorem \ref{weaksp_thm}, we can obtain similar error bound and $l_r$ ball constraint for the proposed method. We present this result as the following corollary. Moreover, we prove that our constraint (\ref{weaksp_def}) is more relaxed than the $l_r$ ball constraint and thus more general.

\begin{corollary}
	Consider weighted $l_1$-LS estimator in (\ref{eq:Mest}) with true parameter $\bbeta^*$ within the $l_r$ ball constraint: $\mathbb{B}_r(R):=\lbk \bbeta^*: \sum_{j=1}^q|\beta_j^*|^r \leq R \rbk$. Assume Assumption \ref{stable_ass} and \ref{weaksp_ass} hold. Further set $w_1=\min\{w_{l,ss'}\}$, $w_2=\max\{w_{l,ss'}\}$, $r_w=w_2/w_1$, $\lambda_N= 4w_1^{-1}\bQ\sqrt{(\log p + 2\log m)/N}$ and $\eta=\lambda_N/\alpha$. Then there exist constants $b_i>0$, such that for any $N \gtrsim(1+r_w)^2|J_{\eta}| \max\{\omega^2,1\}(\log p+2\log m)$, with at least probability:
	$$1-b_1\exp(-b_2N\min\{\omega^{-2},1\})-b_3\exp(-b_4(\log p + 2\log m)),$$
	the estimation error is bounded as follows:
	\begin{equation}
	\|\hbeta-\bbeta^*\|_2  \leq \frac{w_1+2w_2+2\sqrt{w_2}}{\alpha^{\frac{2-r}{2}}}R^{\frac{1}{2}}\lambda_N^{\frac{2-r}{2}} + \frac{4w_2\max\{\omega,1\}}{\bQ\alpha^{1-r}}R\lambda_N^{{2-r}}.
	\label{eq:coro}
	\end{equation}
	\label{weaksp_coro}
\end{corollary}
\vspace{-1.3cm}
\paragraph*{Remark:} (a) The above corollary implies $\alpha^{\frac{r-2}{2}}R^{\frac{1}{2}}\lambda_N^{\frac{2-r}{2}}=o(1)$ and $\alpha^{r-1}R\lambda_N^{{2-r}} / Q = o(1)$ are required to obtain the estimation consistency in the $l_r$ ball setting. After plugging in the choice of $\lambda_N$, we obtain the following $l_r$ ball constraint for $l_1$ regularized estimation of VAR:
	\begin{align}
	&\alpha^{r-2}\bQ^{2-r}R\lsk\frac{\log p + 2\log m}{N}\rsk^{\frac{2-r}{2}}=o(1),\nonumber   \quad \text{and} \\  &\max\{\omega,1\}\alpha^{r-1}\bQ^{1-r}R\lsk\frac{\log p + 2\log m}{N}\rsk^{\frac{2-r}{2}}=o(1).
	\label{eq:lrball_consist}
	\end{align}
In the supplemental material, we proved this constraint is more strict than our weak sparsity constraint (\ref{weaksp_def}). 

(b) It is also worth noting that, in the special case when $\alpha$ is bounded away from zero and $\bQ$ and $\omega$ are bounded away from infinity, the second term in (\ref{eq:coro}) is of higher order than the first term. Thus the convergence rate becomes $\|\hbeta-\bbeta^*\|_2=O\lsk R^{\frac{1}{2}}\lsk(\log p+2\log m)/N \rsk^{1/2-r/4}\rsk=O\lsk R^{\frac{1}{2}} \lsk \log q/N \rsk^{1/2-r/4} \rsk$ with $q=pm^2$ being the number of parameters. This rate is the same as that in regression of independent data \citep{Raskutti2011,negahban2009unified}. 

\section{Simulation Studies}
\label{sec:simu}
In this section, we first describe the implementation of the proposed weighted $l_1$ LS approach (\ref{eq:wls}). Then we present several simulation studies which compare the proposed method with four existing penalized estimations of high-dimensional VAR, which include LASSO \citep{basu2015}, SCAD and MCP \citep{zhu2020nonconcave} and spaLASSO \citep{schweinberger2017high}. Three different VAR order ($p=1,2,3$) and three different sparse scenarios are considered. It turns out in all settings the proposed method achieves substantial improvement over the four existing methods in parameter estimation, network detection and out-of-sample forecast.
\subsection{Practical Implementation}
\label{subsec:opti}
The objective function in the minimization problem (\ref{eq:wls}) can be decomposed as a sum of independent objectives:
$$\sum_{i=1}^m\lmk\frac{1}{N}\norm{\bY_i-\bX \bB_i}_2^2 +\lambda_N\Omega_i(\bB_i)\rmk,$$
where $\bY_i$ and $\bB_i$ are the $i$th column of matrices $\bY$ and $\bB$ respectively, and $\Omega_i(\bB_i)=\sum_{l=1}^p\sum_{j=1}^m w_{l,ss'}|\Phi_{l,ss'}|$. Therefore, the optimization (\ref{eq:wls}) can be solved in parallel by solving the following sub-objectives:
\begin{equation}
\min_{\bB_i}\frac{1}{N}\norm{\bY_i-\bX \bB_i}^2 +\lambda_N\Omega_i(\bB_i), \quad i=1,\cdots,m.
\label{eq:subwls}
\end{equation}
By defining $\tPhi_{l,ss'}=w_{l,ss'}\Phi_{l,ss'}$, $\tB=[\tPhi_1,\cdots,\tPhi_p]'$ and correspondingly $\tX^{(i)}=\lmk \tX^{(i)}_1,\cdots,\tX^{(i)}_{mp} \rmk$ whose $j$th column is $\tX^{(i)}_j=\bX_j\circ w^{(i)}$ with\\ $w^{(i)}=(1/w_{1,i1},\cdots,1/w_{1,im},\cdots,1/w_{p,i1},\cdots,1/w_{p,im})'$, objective (\ref{eq:subwls}) is transformed into a LASSO optimization:
\begin{equation*}
\min_{\tB_i}\frac{1}{N}\|\bY_i-\tX^{(i)} \tB_i\|_2^2 +\lambda_N\|\tB_i\|_1, \quad i=1,\cdots,m,
\label{eq:subwls2}
\end{equation*}
which can be easily solved by existing LASSO algorithms.

In practice, we need to select the VAR order $p$, the penalty parameter $\lambda_N$ and the universal constant $c_i$ in the penalty weights (\ref{eq:weight}). The parameter selection can follow the forward cross-validation approach which is commonly used in high-dimensional VAR model estimation \citep{banbura2010large,song2011large,nicholson2016} and it provides good performance for finite sample as shown in the following simulation studies and real data analysis. 
Firstly, we separate data into two sets:  training dataset $\{1,\cdots,T_0\}$ and validation dataset $\{T_0+1,\cdots,T\}$. Here $T_0$ is pre-specified such as $T_0=\lfloor 0.6T \rfloor$. Then we specify the potential values of $p$ and $c$ such as $p\in\{1,\cdots,4\}$ and $c_i\in\{0.5,5,10,15,20,25,30\}$. For each given pair of $(p,c_i)$, we follow \cite{JSSv033i01} to perform the grid search of $\lambda_N$, which starts from $\lambda_N^{max}$, the smallest value that shrinks all parameters to zero, and then decreases in log linear increments until the value of $\lambda_N^{max}/1000$ is reached. We take 30 values along this grid, and obtain  $4\times7\times30$ triples of $(p,c_i,\lambda_N)$. For each triple of $(p,c_i,\lambda_N)$, we optimize (\ref{eq:wls}) using the training dataset and then calculate 1-step-ahead forecast $\hat{X}_{t+1}^{(p,c_i,\lambda_N)}$ for the validation dataset ($t=T_0,\cdots,T-1$). After that we select the values of $(p,c_i,\lambda_N)=(p^{opt},c_i^{opt},\lambda_N^{opt})$ by minimizing the following Root Mean Squared Forecast Error (RMSFE):
$$RMSFE=\sqrt{\frac{1}{T-T_0}\sum_{t=T_0}^{T-1}\frac{1}{m}\norm{\hat{X}_{t+1}^{(p,c_i,\lambda_N)}-X_{t+1}}_2^2}.$$
Finally, we optimize (\ref{eq:wls}) based on selected  $(p^{opt},c_i^{opt},\lambda_N^{opt})$ and data till $T$.

\subsection{Simulation Setting}

In each study, we simulate the VAR process 100 times and each simulated process has 150 observations. The last 80 points ($t=71,\cdots,150$) is preserved as test dataset for out-of-sample forecast comparison. For LASSO, SCAD, MCP and the proposed method, we apply the aforementioned forward cross-validation to select the tuning parameters, and set data within $t=1,\cdots,40$ as training dataset and data within $t=41,\cdots,70$ as the validation dataset. For spaLASSO, we directly use the code in the online supplemental materials of \cite{schweinberger2017high} to carry out model estimation and prediction. This method uses stability selection \citep{meinshausen2010stability} to sidestep the selection of tuning parameters. Two weight functions are considered in the proposed method:
$$ \text{WLASSO1: }w_{l,ss'}^{(1)}= \exp \left( c_1 \frac{l \, d_{ss'}}{p\,d_{max}} \right); \quad\; \text{WLASSO2: } w_{l,ss'}^{(2)} = \left(1+ \frac{l \, d_{ss'}}{p\,d_{max}}\right)^{c_2}.$$
We consider the following criteria to compare method performance:\vspace{-0.3cm}
\begin{itemize}
	\item $l_1$ estimation error: $\|\hbeta-\bbeta^*\|_1=\underset{l,s,s'}{\sum}|\hat{\Phi}_{ss',l}-\Phi^*_{ss',l} |$.
	\vspace{-0.2cm}
	\item $l_2$ estimation error: $\|\hbeta-\bbeta^*\|_2=\sqrt{\underset{l,s,s'}{\sum}|\hat{\Phi}_{ss',l}-\Phi^*_{ss',l} |^2}$. 
	\vspace{-0.2cm}
	\item Percentage of false zeros: $\text{PFZ}=\underset{l,s,s'}{\sum} I(\hat{\Phi}_{ss',l}=0, \Phi^*_{ss',l}\neq 0)/m^2p $. 
	\vspace{-0.2cm}
	\item Percentage of false nonzeros: $\text{PFNZ}=\underset{l,s,s'}{\sum} I(\hat{\Phi}_{ss',l}\neq 0, \Phi^*_{ss',l}= 0)/m^2p$.
	\vspace{-1.2cm}
	\item RMSFE for $h$-step out-of-sample forecast with $h=1,\cdots,5$.
	\vspace{-0.2cm}
\end{itemize}
To simply the presentation of results, we treat LASSO as benchmark and report the ratio of each method over LASSO. Ratio less than one means the method outperforms LASSO.

\subsection{Simulation of VAR(1)}
\label{subsec:simuVAR1}

First we construct $21\times 21$ lattice with coordinates $\{(x_i,y_j)\}_{i,j=1}^{20}$ as $x_i=0.05i+\delta_i$ and $y_i=0.05i+\delta_{i}$, where $\delta_i$ and $\delta'_{i}$ are independently generated from unif(-0.01,0.01). Then we consider two settings:
\begin{itemize}
	\item Setting 1 randomly selects 100 sites from all 441 vertices in the lattice.\vspace{-0.2cm}
	\item Setting 2 randomly selects 100 sites from the lower left corner ($x_i<0.5$ and $y_i<0.5$) and the upper right corner ($x_i>0.5$ and $y_i>0.5$). 
\end{itemize}
In each setting, we set $\Sigma=0.01I$ and consider three sparse scenarios: 
\begin{itemize}
	\item[(a)] Exactly sparse:  First generate $|\tilde{\Phi}^*_{ss'}|\sim \text{unif}(0.1,0.5)$ then set  $|\Phi^*_{ss'}|=|\tilde{\Phi}^*_{ss'}|I(d_{ss'}\leq d_0)$ with $d_0=0.05$ (setting 1) or $d_0=0.06$ (setting 2).\vspace{-0.2cm}
	\item[(b)] Weakly sparse (fast decay): $|\Phi^*_{ss'}|=0.55/\exp(20\,d_{ss'})$.\vspace{-0.2cm}
	\item[(c)] Weakly sparse (slow decay): $|\Phi^*_{ss'}|=0.25/\exp(5\,d_{ss'})$.\vspace{-0.2cm}
\end{itemize}
The sign of $\Phi^*_{ss'}$ is randomly selected. Scenario (a) stands for exact sparsity and scenario (b) and (c) are for weak sparsity. $|\Phi^*_{ss'}|$ in scenario (b) decays much faster than that in scenario (c), thus the transition matrix in (b) is more sparsifiable. 
To guarantee the VAR(1) process is stationary, the above generation procedure is repeated until all eigenvalues of $\Phi^*$ are within (-1,1). Figure \ref{fig:simu_sites} in the supplemental material shows the generated spatial locations. 

\paragraph{Simulation Results of VAR(1)}

The empirical results for setting 1 and 2 are very similar. Thus, we mainly focus on setting 1. Table \ref{table:simu1} in the supplemental material lists the performance of different methods. 
In terms of model fitting, the proposed method achieves considerable improvement over the other four methods in all three scenarios, highlighting the advantage of incorporating spatial and temporal information. The only exception is PFNZ, in which the proposed method is out-performed by MCP and spaLASSO. This is because MCP and spaLASSO are too conservative and severely underestimate nonzero parameters, thus their PFNZ are low but their PFZ are very high. In particular, in scenario (a), the PFZ and PFNZ of the proposed method are only 5\% and 20\% of those from LASSO, and the $l_1$ and $l_2$ estimation errors are reduced by around 60\% compared with LASSO.  In contrast, SCAD, MCP and spaLASSO do not outperform LASSO in the three scenarios and suffer from underestimation of nonzero parameters. 
Figure \ref{fig:net_ill} in Section 1 depicts the network detection results of one randomly selected replicate in scenario (a) and the results are consistent with what we observed in PFZ and PFNZ: the proposed method performs the best and provides desirable network estimation, while the other four methods severely underestimates true connections while LASSO also overestimates wrong connections. 

Figure \ref{fig:forecast1} in the supplemental material plots the RMSFE ratio between each method and the benchmark (LASSO). We can see the proposed method significantly improves over LASSO at $h=1,2,3,4$ in all scenarios and at $h=5$ in scenario (b). In contrast, the other three methods do not show obvious advantages over LASSO and sometimes are even worse due to their severe underestimation of nonzero parameters. In addition, the performance of WLASSO1 and WLASSO2 are very close, which means the proposed method is not sensitive to the choice of weight function. The following simulation studies of VAR(2) and VAR(3) and the real data analysis also confirm this robustness. 

\subsection{Simulation for VAR(2) and VAR(3)}

The detailed simulation settings are reported in the supplemental material. Similar to the results of VAR(1), The proposed method also demonstrate clear advantages over other approaches in model fitting, network detection and out-of-sample forecasting (Table \ref{table:simu_fitp2}-\ref{table:simu_fitp3} and Figure \ref{fig:net_p2}-\ref{fig:forecast_p23}). Moreover, as shown in Figure \ref{fig:forecast_p23}, the improvement over LASSO on forecasting becomes more obvious as $p$ increases. This is because our method penalizes parameters according to not only spatial distance but also temporal lags.

\section{Traffic Data Analysis}
\label{sec:realdat}
The real data contains the hourly traffic volumes recorded on 79 sites on highways around Des Moines, Iowa. The records are hourly data from 2014-09-20 to 2014-11-02 (six weeks and two days), with a total of 1056 observations for each site. These 79 sites are shown in Figure \ref{fig:site} in the the supplemental material.  

For each site $i$, the volume series $\{ z_{st}\}$ ($s=1,\cdots,79; \; t=1,\cdots,1056$) has strong weekly periodicity, i.e. its weekly trend is repeated every 168 time points. For each time point $t$, we use $d=t\,\text{mod}\,(168)$ to denote the hour of the time point $t$ in one week. We model the volume series $\{ z_{st}\}$ as follows:
\begin{align}
&z_{st} = \mu_{sd} + \sigma_{sd} x_{st}, \quad E(x_{st}) = 0 \text{ and } E(x_{st}^2)=1, \quad \log(\sigma_{sd}) = a_s + b_s \log(\mu_{sd}),\nonumber\\
&X_t=(x_{1t},\cdots,x_{mt}), \quad X_t=\Phi_1X_{t-1}+\cdots+\Phi_pX_{t-p}+\varepsilon_t.
\label{eq:stage2}
\end{align}
Here $\{\mu_{id}\}_{d=1}^{168}$ is the weekly trend of $\{ z_{it}\}$, and $\{x_{it}\}_{t=1}^{1056}$ is the series after subtracting the trend and standardization, which is assumed to be stationary.  $E(x_{it}) = 0$ and $E(x^2_{it})=1$ guarantee $\sigma_{id}$ and $x_{it}$ are identifiable. The following two-stage procedure is carried out for estimation and forecasting. 

\paragraph*{Stage 1: Estimate $\boldsymbol{\mu_{id}}$, $\boldsymbol{\sigma_{id}}$ and series $\boldsymbol{\{x_{it}\}}$} We first use the local linear kernel regression \citep{fan1995local} to estimate $\{\mu_{id}\}_{d=1}^{168}$, and obtain detrended series $y_{it}:=  z_{it} - \hat{\mu}_{id}$. Since we have multiple $y_{it}$'s at each $d$, we can approximate $\sigma_{id}$ by the standard error of these $y_{it}$'s (i.e. $\hat{\sigma}_{id}$ is the standard error of $\{y_{it}: t\,\text{mod}\,(168)=d\}$). Then we regress $\log(\hat{\sigma}_{id})$ on $\log(\hat{\mu}_{id})$ to estimate $a_i$ and $b_i$. Finally, the estimate of series  $\{x_{it}\}$ can be obtained by $\hat{x}_{it}=(z_{it}-\hat{\mu}_{id})/\exp(\hat{a}_i+\hat{b}_i\log(\hat{\mu}_{id}))$. Figure \ref{fig:reg} in the supplemental material illustrates the result of one site in Stage 1.

Notice that some stretch of observations in $\{z_{it}\}$ are zero. This may be a result from road construction or maintenance at that time. These zero observations are considered as outliers and excluded when estimating $\{\mu_{id}\}$ and $\{\sigma_{id}\}$. The following procedure is applied for outlier screening. For a given $d$, we have six to seven $z_{it}$'s. If the median of these $z_{it}$'s is above 30, but one of them, say $z_{it_0}$, is zero, we mark $z_{it_0}$ as outlier. In addition, we used the idea of boxplot to identify outliers: if $z_{it_0}$ is below  the interquantile of 25\% quantile or above the interquantile of 75\% quantile, $z_{it_0}$ is marked as outlier. We exclude these outliers when estimating $\{\mu_{id}\}$ and $\{\sigma_{id}\}$, but attribute them to component $\{x_{it}\}$.

\paragraph*{Stage 2: Modeling $\boldsymbol{\{\hat{X}_t\}}$} Set $\hat{X}_t=(\hat{x}_{1t}, \cdots \hat{x}_{mt})'$, we apply VAR, LASSO and the proposed method to estimate model (\ref{eq:stage2}) and carry out forecasting. Here we divide the time period into four sub-periods: (1) weekday peak time (6am - 8pm); (2) weekday off-peak time (9pm - next day 5am); (3) weekend peak time (8am - 8pm); (4) weekend off-peak time (9pm - next day 7am). We carried out 1 to 4 steps ahead forecasting for the last two weeks. To incorporate the spatial location information, we calculate the road distances among the 79 sites. If there is a highway path from site $i$ to site $j$, $d_{ss'}$ is the road distance of this path, otherwise we set $d_{ss'}=d_{max}$ where $d_{max}:=\max\{d_{ss'}: \text{there is a road path from } i \text{ to } j\}$. The following four kinds of weight functions are considered:
\begin{align*}
	&\text{WLASSO1:}\;  w^{(1)}_{l,ss'}=\exp\lsk c_1\frac{l\,d_{ss'}}{p\,d_{max}} \rsk, \quad \text{WLASSO2:}\; w^{(2)}_{l,ss'}=\lsk1+\frac{l\,d_{ss'}}{p\,d_{max}} \rsk^{c_2}, \\
	&\text{WLASSO3:}\; w^{(3)}_{l,ss'}=\lsk \frac{l}{p}\exp\lsk\,\frac{d_{ss'}}{d_{max}}\rsk\rsk^{c_3}, \quad \text{WLASSO4:}\; w^{(4)}_{l,ss'}=\exp\lsk c_4\frac{d_{ss'}}{d_{max}}\rsk.
\end{align*}
We also tried another setting in which $d_{ss'}=\infty$ if there is no road path between site $i$ and site $j$. This setting forces the corresponding $\Phi_{ss',l}$ to be zero. In practice, these two distance settings provide very similar network detection and forecasting performance. For both LASSO and the proposed method, VAR order $p$ is selected from $\{1,\cdots,6\}$. 
Table \ref{table:nobs} in the supplemental material lists the partition of training dataset, validation dataset and test dataset. In short, the last two weeks are the test data, the last third and forth weeks are the validation data. It turns out the performance of WLASSO1, WLASSO2 and WLASSO3 are very close and WLASSO4 behaves slightly worse, thus we only report the result of WLASSO1. 

\paragraph*{Summary of Fitting and Forecasting Results} 
Table \ref{table:real_order} in the supplemental material lists the selected orders of LASSO and WLASSO1 through forward cross-validation. For VAR without any penalty, we fix $p=1$ which gives the best forecast. WLASSO1 selects $p$ as 1 or 2 for all sub-periods, but LASSO selects $p=5$ for weekend peak time. $p=5$ means one site may be influenced by another site even after five hours, which seems to be unreasonable. This fallacy is because LASSO penalizes parameters equally no matter what the temporal lag is. The forecasting RMSFEs are listed in Table \ref{table:forecast} in the supplemental material. Unsurprisingly LASSO and WLASSO1 behave much better than VAR. Meanwhile WLASSO1 is superior than LASSO for all scenarios except weekday peak time with $h=1$. In particular for weekend peak time, WLASSO1 outperforms LASSO by reducing RMSFE by 17\%, 9\%, 8\% and 6\% for $h$=1, 2, 3 and 4 respectively. It also reduces RMSFE by 8\% in weekend off-peak time with $h=1$. To examine the significance of such improvements, we carried out Diebold-Mariano (DM) test \citep{diebold2002comparing} for each sub-period. The test results state that WLASSO1 is significantly better than LASSO in weekend peak time. 

In addition, WLASSO1 gives more reasonable network estimation than LASSO does in all sub-periods. For instance, Figure \ref{fig:weekendpeak} in the supplemental material displays the network estimation in weekend peak time by LASSO and WLASSO1 respectively. LASSO connects some sites far from each other or even in the opposite directions, which is counter-intuitive, while WLASSO1 only connects the sites close to each other. One may argue that it is unnatural to have dependence only within close sites, and two sites can still have similarities even if they are far from each others. For example, they may have peaks at around 8AM and 5PM on weekday. However, such similarities are in the weekly trend of each series, thus the spatial dependence among de-trended series only exists within close sites. 

\section{Conclusion and Discussion}
\label{sec:conclusion}

In this paper, we introduced a data-driven weighted $l_1$ regularized estimation of high-dimensional VAR model for spatio-temporal data. This method incorporates spatial distance and temporal lags to construct penalty weights. Its optimization is straightforward and easy to implement by existing algorithms. Its theoretical properties has been explored in both exactly sparse scenario and weakly sparse scenario, as well as the conditions for consistency, which indicates the proposed method achieves smaller error bounds than LASSO. The theoretical results of $l_1$ regularization in weakly sparse scenario are new and have not been addressed in the time series framework. Our definition of weak sparsity is also more general than the $l_r$ ball setting in the literature. To evaluate the model performance, we compare the proposed method with four existing penalized VAR estimation through simulation studies, which demonstrates the proposed method can obtain more reasonable network detection and substantial improvement on model fitting and forecasting. Real application on a traffic dataset also indicates advantages of the proposed method over LASSO.

Adaptive LASSO \citep{zou2006adaptive, wang2007autoregression} can be seen as a variant of the proposed methodology. However, the idea and rationale of adaptive LASSO is different from the proposed method. Specifically, in the adaptive LASSO, the penalty weights are determined by an initial $\sqrt{T}$-consistent estimator without considering any spatial and temporal information. In practice, the initial estimator may not be precise enough to provide proper weights. In contrast, our proposed method constructs the weights based on the spatial and temporal information of the data instead of any initial estimator, and its sense of ``adaptive" lies in data-driven selection of certain constants in the weight functions. Thus, the proposed method not only preserves the spatio-temporal structure of the data but also avoids the impact of any possible inaccuracy and variation of the initial estimator. 

In this paper, the tuning parameters are selected by forward cross-validation and it yields reasonable performances as reported in the numerical analysis. Another popular approach in the literature is the BIC criterion \citep{band, wang2007autoregression, wang2007tuning} or high-dimensional BIC (HBIC) criterion \cite{zhu2020nonconcave}. 
However, BIC and HBIC require estimation of covariance matrix $\Sigma$, and the traditional estimation of $\Sigma$ is infeasible in the high-dimensional regime in which the number of observations $T$ is smaller than the number of time series components $m$. In such cases, a feasible solution is to apply penalized estimation for $\Sigma$, but it will involve another tuning parameter selection and is more expensive in computation. The optimal procedure of tuning parameter selection for high-dimensional time series, especially for ultra high-dimensional time series, and the corresponding theoretical properties is out of the scope of this paper, but it is an interesting topic for future study.

\vskip 14pt
\noindent {\large\bf Supplementary Materials}

Supplementary material contains three parts: (1) proofs of theorems, propositions and corollaries; (2) simulation setting of VAR(2) and VAR(3); (3) tables and figures from simulation studies and real data analysis.
\vskip 14pt
\noindent {\large\bf Acknowledgements}

This research was supported by National Science Foundation 1455172, 1934985, 1940124, 1940276, USAID, Xerox PARC Faculty Research Award, Cornell University Atkinson’s Center for a Sustainable Future.
\par

\markboth{\hfill{\footnotesize\rm Zhenzhong Wang, Abolfazl Safikhani, Zhengyuan Zhu and David S.\ Matteson} \hfill}
{\hfill {\footnotesize\rm High-dimensional Spatio-temporal VAR} \hfill}

\bibhang=1.7pc
\bibsep=2pt
\fontsize{9}{14pt plus.8pt minus .6pt}\selectfont
\renewcommand\bibname{\large \bf References}
\expandafter\ifx\csname
natexlab\endcsname\relax\def\natexlab#1{#1}\fi
\expandafter\ifx\csname url\endcsname\relax
  \def\url#1{\texttt{#1}}\fi
\expandafter\ifx\csname urlprefix\endcsname\relax\def\urlprefix{URL}\fi

\bibliographystyle{chicago}      
{\linespread{1}\selectfont\bibliography{wlasso}}   

\vskip .3cm
\noindent
Department of Statistics, Iowa State University, Ames, IA.
\noindent
E-mail: zwang1@iastate.edu
\vskip -0.2cm
\noindent
Department of Statistics, University of Florida, Gainesville, FL.
\noindent
E-mail: a.safikhani@ufl.edu
\vskip -0.2cm
\noindent
Department of Statistics, Iowa State University, Ames, IA.
\noindent
E-mail: zhuz@iastate.edu
\vskip -0.2cm
\noindent
Department of Statistics and Data Science, Cornell University, Ithaca, NY.
\noindent
E-mail: matteson@cornell.edu

\newpage
\fontsize{12}{14pt plus.8pt minus .6pt}\selectfont \vspace{0.8pc}
\centerline{\large\bf Regularized Estimation in High-Dimensional Vector}
\vspace{2pt} \centerline{\large\bf Auto-Regressive Models using Spatio-Temporal Information}
\vspace{.4cm} \centerline{Zhenzhong Wang, Abolfazl Safikhani, Zhengyuan Zhu and David S.\ Matteson} \vspace{.4cm} \centerline{\it
	Iowa State University, University of Florida and Cornell University} \vspace{.55cm} \fontsize{9}{11.5pt plus.8pt minus
	.6pt}\selectfont

\setcounter{table}{0}
\setcounter{figure}{0}
\setcounter{section}{0}
\section{Detailed Proofs}

\subsection{Proof of Theorem 1}
\begin{proof}
	Recall that the RE condition (\ref{eq:RE}) and Derivation (\ref{eq:der}) condition hold with probability $1-b_1\exp(-b_2 N\min\{\omega^{-2},1\})-b_3\exp(-b_4(\log p + 2\log m))$:
	\begin{align}
	\text{Restricted Eigenvalue (RE):} \quad &\theta'\hGamma\theta\geq \alpha\norm{\theta}_2^2 - \tau\norm{\theta}_1^2, \quad \forall \theta\in R^q, \\
	\text{Derivation condition:}\quad & \|\hgamma-\hGamma\bbeta^*\|_{\infty}\leq\bQ\sqrt{\frac{\log p + 2\log m}{N}}.
	\end{align}
	In the following proof, we firstly assume the RE condition (\ref{eq:RE}) and Derivation (\ref{eq:der}) hold, then derive the upper bounds of estimation errors in Theorem 1. Since these two conditions hold with high probability, we can conclude that the upper bounds hold with the same probability. 

Based on the definition of our proposed method, we have
	$$-2\hbeta'\hgamma + \hbeta'\hGamma\hbeta +\lambda_N\Omega(\hbeta)\leq -2{\bbeta^*}'\hgamma + {\bbeta^*}'\hGamma\bbeta^* +\lambda_N\Omega(\bbeta^*)$$
	Set $\bv=\hbeta-\bbeta^*$, since $\bbeta^*=\bbeta^*_J$, we can obtain:
	\begin{eqnarray}
	\bv'\hGamma\bv 
	&\leq & 2\bv'(\hgamma-\hGamma\bbeta^*) + \lambda_N\lbk\Omega(\bbeta^*) - \Omega(\bbeta^*+\bv)\rbk \nonumber \\
	&\leq & 2\norm{\bv}_1\|\hgamma-\hGamma\bbeta^*\|_{\infty} + \lambda_N\lbk\Omega(\bbeta^*) - \Omega(\bbeta^*+\bv)\rbk \nonumber \\
	& = & 2\norm{\bv}_1\|\hgamma-\hGamma\bbeta^*\|_{\infty} + \lambda_N\lbk\Omega(\bbeta^*_J) - \Omega(\bbeta^*_J+\bv_J) - \Omega(\bv_{J^C}) \rbk \nonumber\\
	&\leq &  2\norm{\bv}_1\|\hgamma-\hGamma\bbeta^*\|_{\infty} + \lambda_N\lbk\Omega(\bv_J) - \Omega(\bv_{J^C}) \rbk. 
	\label{proof:exact_upb1}
	\end{eqnarray}
	Suppose Derivation condition (\ref{eq:der}) hold, since $\lambda_N\geq4\bQ\sqrt{(\log p + 2\log m)/N}$, we can upper bound $2\norm{\bv}_1\|\hgamma-\hGamma\bbeta^*\|_{\infty}$ by
	$$2\norm{\bv}_1\|\hgamma-\hGamma\bbeta^*\|_{\infty} \leq 2\bQ\sqrt{\frac{\log p + 2\log m}{N}}\norm{\bv}_1 \leq \frac{\lambda_N}{2}\norm{\bv}_1.$$
	Coupled with $\min\{w_{l,ss'}:(l,ss')\in J^C\}=1$ and $r_w=\max\{w_{l,ss'}:(l,ss')\in J\}$, the upper bound in (\ref{proof:exact_upb1}) becomes:
	\begin{eqnarray}
	\bv'\hGamma\bv 
	&\leq & \lambda_N\lbk \frac{1}{2}\norm{\bv}_1 + r_w\norm{\bv_J}_1 - \norm{\bv_{J^C}}_1  \rbk \nonumber\\
	&\leq & \frac{1+2r_w}{2}\lambda_N\norm{\bv_J}_1 \leq  \frac{1+2r_w}{2}\lambda_N\sqrt{k}\norm{\bv_J}_2 \nonumber\\
	&\leq & \frac{1+2r_w}{2}\lambda_N\sqrt{k}\norm{\bv}_2. 
	\label{proof:exact_upb2}
	\end{eqnarray}
	In particular, $\bv'\hGamma\bv\geq 0$ implies $\norm{\bv_{J^C}}_1\leq (1+2r_w)\norm{\bv_{J}}_1$. Thus
	\begin{equation}
	\norm{\bv}_1 \leq (2+2r_w)\norm{\bv_{J}}_1 \leq (2+2r_w)\sqrt{k}\norm{\bv_J}_2 \leq (2+2r_w)\sqrt{k}\norm{\bv}_2.
	\label{eq:norm}
	\end{equation}
	Suppose RE condition (\ref{eq:RE}) holds and $\tau=\alpha\max\{\omega^2,1\}(\log p+\log m)/N$, we have
	\begin{equation}
	\bv'\hGamma\bv\geq \alpha\norm{\bv}_2^2 - \tau\norm{\bv}_1^2\geq \lsk\alpha-4(1+r_w)^2k\tau\rsk\norm{\bv}_2^2.
	\end{equation}
	Moreover, $N \gtrsim (1+r_w)^2\max\{\omega^2,1\}k(\log p+2\log m)$ guarantees $4(1+r_w)^2k\tau\leq\alpha/2$, which indicates $\bv'\hGamma\bv\geq \alpha\norm{\bv}_2^2/2$. Together with (\ref{proof:exact_upb2}) we have $\alpha\norm{\bv}_2^2 \leq (1+2r_w)\lambda_N\sqrt{k}\norm{\bv}_2$, thus the $l_2$ error is bounded by
	\begin{equation}
	\|\bv\|_2 \leq \frac{1+2r_w}{\alpha}\sqrt{k}\lambda_N.
	\label{eq:l1}
	\end{equation}
	Based on (\ref{proof:exact_upb2}) and (\ref{eq:norm}), we can obtain the other two upper bounds w.r.t $l_1$ norm and in-sample prediction:
	\begin{equation}
	\|\bv\|_1 \leq \frac{2+6r_w+4r_w^2}{\alpha}k\lambda_N,
	\label{eq:l2}
	\end{equation}
	\begin{equation}
	\bv'\hGamma\bv \leq \frac{(1+2r_w)^2}{2\alpha}k\lambda_N^2.
	\label{eq:b3}
	\end{equation}
	For the number of false zero, we have
	\begin{eqnarray}
	\left| \mathrm{supp}(\bbeta^*)\backslash \mathrm{supp}(\hbeta)\right| &=& \sum_{j\in J,\hat{\beta}_j=0}1 \leq \sum_{j\in J,\hat{\beta}_j=0}\frac{\beta^*_j}{s_0} = \sum_{j\in J,\hat{\beta}_j=0}\frac{|v_j|}{s_0} \nonumber\\ 
	&\leq&\frac{\|\bv\|_1}{s_0} \leq\frac{2+6r_w+4r_w^2}{s_0\alpha}k\lambda_N.
	\label{eq:fz}
	\end{eqnarray}
	
	Moreover, if we set $\tilde{\bbeta}:=\{\hat{\beta}_{j}I(\hat{\beta}_j>\lambda_N)\}$, then the number of false non-zero of $\tilde{\bbeta}$ is bounded by
	$$\left| \mathrm{supp}(\tilde{\bbeta})\backslash \mathrm{supp}(\bbeta^*)\right| = \sum_{j\in J^C}I(|\hat{\beta}_j|>\lambda_N) \leq  \sum_{j\in J^C}\frac{|\hat{\beta}_j|}{\lambda_N}
	 = \frac{1}{\lambda_N} \|\bv_{J^C}\|_1.$$
	Since  $\norm{\bv_{J^C}}_1\leq (1+2r_w)\norm{\bv_{J}}_1$, we have $\norm{\bv_{J^C}}_1\leq \frac{1+2r_w}{2(1+r_w)}\|\bv\|_1$ and thus
	\begin{equation}
	\left| \mathrm{supp}(\tilde{\bbeta})\backslash \mathrm{supp}(\bbeta^*)\right| \leq \frac{1+2r_w}{2(1+r_w)}\frac{\|\bv\|_1}{\lambda_N}\leq(1+2r_w)^2\frac{k}{\alpha}.
	\label{eq:fnz}
	\end{equation}
	At last, since $N \gtrsim (1+r_w)^2\max\{\omega^2,1\}k(\log p+2\log m) \gtrsim \max\{\omega^2,1\}(\log p+2\log m)$, based on Proposition (4.2) and (4.3) in Basu and Michailidis (2015) the RE condition (\ref{eq:RE}) and Derivation condition (\ref{eq:der}) holds with probability at least $1-b_1\exp(-b_2N\min\{\omega^{-2},1\})-b_3\exp(-b_4(\log p + 2\log m))$. Thus the five upper bounds (\ref{eq:l1}) to (\ref{eq:fnz}) hold with the same probability.
\end{proof}

\subsection{Proof of Theorem 2}
We still set $\bv=\hbeta-\bbeta^*$ and suppose RE condition (\ref{eq:RE}) and Derivation (\ref{eq:der}) hold. By triangular inequality, we have

\begin{eqnarray*}
	\Omega(\bbeta^*) - \Omega(\bbeta^*+\bv) &=& \Omega(\bbeta^*_{J_\eta}) + \Omega(\bbeta^*_{J^C_\eta}) - \Omega(\bbeta^*_{J_\eta}+\bv_{J_\eta}) - \Omega(\bbeta^*_{J^C_\eta}+\bv_{J^C_\eta}) \nonumber \\
	& \leq & \Omega(\bbeta^*_{J_\eta}) + \Omega(\bbeta^*_{J^C_\eta}) - \Omega(\bbeta^*_{J_\eta}+\bv_{J_\eta}) - \Omega(\bv_{J^C_\eta}) + \Omega(\bbeta^*_{J^C_\eta}) \nonumber \\
	& \leq & \Omega(\bv_{J_\eta}) - \Omega(\bv_{J^C_\eta}) + 2\Omega(\bbeta^*_{J^C_\eta}) \nonumber \\
	& \leq & w_2(\eta) \|\bv_{J_\eta}\|_1 - w_1(\eta) \|\bv_{J^C_\eta}\|_1 + 2w_2(\eta) \|\bbeta^*_{J^C_\eta}\|_1.
\end{eqnarray*}

Utilizing the above result and $\lambda_N = 4w^{-1}_1(\eta)\bQ\sqrt{(\log p + 2\log m)/N} \geq 4w^{-1}_1(\eta)\|\hgamma-\hGamma\bbeta^*\|_{\infty}$, we have
\begin{eqnarray}
\bv'\hGamma\bv 
&\leq & 2\bv'(\hgamma-\hGamma\bbeta^*) + \lambda_N\lbk\Omega(\bbeta^*) - \Omega(\bbeta^*+\bv)\rbk \nonumber \\
&\leq & 2\norm{\bv}_1\|\hgamma-\hGamma\bbeta^*\|_{\infty} + \lambda_N\lbk w_2(\eta) \|\bv_{J_\eta}\|_1 - w_1(\eta) \|\bv_{J^C_\eta}\|_1 + 2w_2(\eta) \|\bbeta^*_{J^C_\eta}\|_1 \rbk \nonumber\\
&\leq &  \frac{w_1(\eta)}{2}\lambda_N \norm{\bv}_1 +  \lambda_N\lbk w_2(\eta) \|\bv_{J_\eta}\|_1 - w_1(\eta) \|\bv_{J^C_\eta}\|_1 + 2w_2(\eta) \|\bbeta^*_{J^C_\eta}\|_1 \rbk \nonumber\\
&= & \lambda_N\lbk \frac{w_1(\eta)+2w_2(\eta)}{2} \|\bv_{J_\eta}\|_1 - \frac{w_1(\eta)}{2} \|\bv_{J^C_\eta}\|_1 + 2w_2(\eta) \|\bbeta^*_{J^C_\eta}\|_1 \rbk. 
\label{proof:weak_upb1}
\end{eqnarray}
The above inequality and $\bv'\hGamma\bv \geq 0$ implies $\|\bv_{J^C_\eta}\|_1 \leq (1+2r_w(\eta))\|\bv_{J_\eta}\|_1 + 4r_w(\eta)\|\bbeta^*_{J^C_\eta}\|_1$, which gives an inequality between $\|\bv\|_1$ and $\|\bv\|_2$ by
\begin{eqnarray}
\|\bv\|_1 & = &\|\bv_{J_{\eta}}\|_1+\|\bv_{J_{\eta}^{C}}\|_1 \nonumber\\
&\leq &(2+2r_w(\eta))\|\bv_{J_{\eta}}\|_1+4r_w(\eta)\|\bbeta_{J^C_{\eta}}^*\|_1 \nonumber\\
&\leq & (2+2r_w(\eta))\sqrt[]{|J_{\eta}|}\|\bv_{J_{\eta}}\|_2+4r_w(\eta)\|\bbeta_{J^C_{\eta}}^{*}\|_1 \nonumber\\
&\leq & (2+2r_w(\eta))\sqrt{|J_{\eta}|}\|\bv \|_2+4r_w(\eta)\|\bbeta_{J^C_{\eta}}^{*}\|_1. 
\label{proof:weak_l1l2}
\end{eqnarray}
Thus we have $\|\bv\|_1^2 \leq 8(1+r_w(\eta))^2|J_{\eta}|\|\bv\|_2^2 + 32r^2_w(\eta)\|\bbeta^*_{J^C_\eta}\|_1^2$. Substitute this into the RE condition (\ref{eq:RE}), and $N \gtrsim (1+r_w(\eta))^2|J_{\eta}|\max\{\omega^2,1\}(\log p+2\log m)$ guarantees $8(1+r_w(\eta))^2\tau |J_{\eta}| \leq \alpha/2$, we have
\begin{align}
\bv'\hGamma\bv &\geq \lsk \alpha  - 8(1+r_w(\eta))^2\tau |J_{\eta}| \rsk \|\bv\|_2^2 - 32r^2_w(\eta)\tau \|\bbeta^*_{J^C_\eta}\|_1^2 \nonumber\\
&\geq  \frac{\alpha}{2} \|\bv\|_2^2 -32r^2_w(\eta)\tau \|\bbeta^*_{J^C_\eta}\|_1^2.
\label{proof:weak_lob}
\end{align}	
inequality (\ref{proof:weak_upb1}) also implies
\begin{align}
\bv'\hGamma\bv &\leq \lambda_N\lbk \frac{w_1(\eta)+2w_2(\eta)}{2}\sqrt{|J_\eta|} \|\bv_{J_\eta}\|_2 + 2w_2(\eta) \|\bbeta^*_{J^C_\eta}\|_1 \rbk \nonumber\\
&\leq \lambda_N\lbk \frac{w_1(\eta)+2w_2(\eta)}{2}\sqrt{|J_\eta|} \|\bv\|_2 + 2w_2(\eta) \|\bbeta^*_{J^C_\eta}\|_1 \rbk.
\label{proof:weak_upb2}
\end{align}
Put the upper bound (\ref{proof:weak_upb2}) and lower bound (\ref{proof:weak_J}) of $\bv'\hGamma\bv$ together, we have
$$\frac{\alpha}{2} \|\bv\|_2^2 - \frac{w_1(\eta)+2w_2(\eta)}{2}\lambda_N\sqrt{|J_\eta|} \|\bv\|_2 - 2w_2(\eta)\lambda_N \|\bbeta^*_{J^C_\eta}\|_1   - 32r^2_w(\eta)\tau \|\bbeta^*_{J^C_\eta}\|_1^2 \leq 0.$$

Solving this quadratic inequality and using $\tau=\alpha\max\{\omega^2,1\}(\log p+\log m)/N= \alpha\max\{\omega^2,1\} w_1^2(\eta)\lambda^2_N/(16\bQ^2)$, we can get
\begin{align}
\|\bv\|_2 & \leq  \frac{1}{\alpha}\lbk \frac{w_1(\eta)+2w_2(\eta)}{2}\sqrt{|J_{\eta}|}\lambda_N \right.\nonumber\\
&+ \left.\sqrt{\frac{(w_1(\eta)+2w_2(\eta))^2}{4}|J_{\eta}|\lambda^2_N + 4w_2(\eta)\alpha\lambda_N \|\bbeta^*_{J^C_\eta}\|_1 + 64r^2_w(\eta)\alpha \tau \|\bbeta^*_{J^C_\eta}\|^2_1 } \rbk \nonumber\\
& \overset{(ii)}\leq   \frac{w_1(\eta)+2w_2(\eta)}{\alpha}\sqrt{|J_{\eta}|}\lambda_N + 2\sqrt{\frac{w_2(\eta)\lambda_N\|\bbeta^*_{J^C_\eta}\|_1}{\alpha}} + \frac{4w_2(\eta)\max\{\omega,1\}}{\bQ} \lambda_N \|\bbeta^*_{J^C_\eta}\|_1\nonumber\\
& \leq   \frac{1+2r_w(\eta)}{\alpha}\sqrt{|J_{\eta}|}\widetilde{\lambda}_N + 2\sqrt{\frac{r_w(\eta)\widetilde{\lambda}_N\|\bbeta^*_{J^C_\eta}\|_1}{\alpha}} + \frac{4r_w(\eta)\max\{\omega,1\}}{\bQ} \widetilde{\lambda}_N \|\bbeta^*_{J^C_\eta}\|_1.
\label{proof:weak_l2}
\end{align}

Here inequality (ii) uses the fact that $\sqrt{a^2+b^2+c^2}\leq a+b+c$ for any positive quantities $a$, $b$ and $c$. The upper bound of $\|\bv\|_1$  and $\bv'\hGamma\bv$ are directly from inequalities (\ref{proof:weak_l1l2}) and (\ref{proof:weak_upb2}).

Since $N \gtrsim (1+r_w(\eta))^2|J_{\eta}|\max\{\omega^2,1\}(\log p+2\log m) \gtrsim \max\{\omega^2,1\}(\log p+2\log m)$, the RE condition (\ref{eq:RE}) and Derivation condition (\ref{eq:der}) holds with probability at least $1-b_1\exp(-b_2N\min\{\omega^{-2},1\})-b_3\exp(-b_4(\log p + 2\log m))$. Thus (\ref{proof:weak_l1l2}), (\ref{proof:weak_upb2}) and (\ref{proof:weak_l2}) hold with the same probability.

Finally, if we plug in the weak sparsity constraint (2.7) into inequality (\ref{proof:weak_l2}), we will obtain that the upper bound of $\|\hbeta-\bbeta^*\|$ is $o(1)$. Thus $\hbeta\xrightarrow{p}\bbeta^*$ as $m,T\rightarrow\infty$.

\subsection{Proof of  Proposition 1}
Suppose the symmetric transition matrix $\Phi$ has real eigenvalues $\lambda_1,\cdots,\lambda_m$ with corresponding real orthonormal eigenvectors $\bp_1,\cdots,\bp_m$, it has spectral decomposition $\Phi=P\Lambda P'$ with $\Lambda=\text{diag}(\lambda_1,\cdots,\lambda_m)$ and $P=[\bp_1,\dots,\bp_m]$. From the definition of spectral radius and spectral norm, it is clear that $\rho(\Phi)=\norm{\Phi}_2=\underset{1\leq i\leq m}{\max} |\lambda_i|<1$. Since $\left| I_m-\Phi z\right|=|P(I_m-z\Lambda)P'|=|P||I_m-z\Lambda||P'|$ and $|P|\neq0$, the roots of $|I_m-\Phi z|=0$ are $\frac{1}{\lambda_i}$ ($i=1,\cdots,m$). By the stationarity of VAR(1) process, we have $|\lambda_i|<1$ for all $i$. 

For $\mu_{\max}(\bPhi)$ and $\mu_{\min}(\bPhi)$, firstly we have 
$$\bPhi^H(z)\bPhi(z)=P(\overline{I_m-\Lambda z})P'P(I-\Lambda z)P=
P
\begin{bmatrix}
|1-\lambda_1z|^2 &  &  \\
& \ddots  &  \\
&  & |1-\lambda_mz|^2
\end{bmatrix}
P,
$$
whose eigenvalues are $|1-\lambda_iz|^2$, $i=1,\cdots,m$. Then
\begin{align*}
\mu_{\max}(\bPhi)&=\underset{|z|=1}{\max}\;\Lambda_{\max}(\bPhi^H(z)\bPhi(z))=\underset{|z|=1}{\max}\,\underset{i}{\max}|1-\lambda_iz|^2\\
&=(1+\max_{i}|\lambda_i|)^2=(1+\rho(\Phi))^2,\\
\mu_{\min}(\bPhi)&=\underset{|z|=1}{\min}\;\Lambda_{\min}(\bPhi^H(z)\bPhi(z))=\underset{|z|=1}{\min}\,\underset{i}{\min}|1-\lambda_iz|^2\\
&=(1-\max_{i}|\lambda_i|)^2=(1-\rho(\Phi))^2.
\end{align*} 
Since $\rho(\Phi)<1$, we have $\mu_{\max}(\bPhi)<4$. Further for VAR(1), $\mu_{\min}(\tilde{\bPhi})$ equals to $\mu_{\min}(\bPhi)$. By the definition of $\alpha$, $\bQ$ and $\omega$, we can conclude that, $\alpha$ is bounded from zero, $\bQ$ and $\omega$ are bounded away from infinity if and only if  $\Lambda_{\max}(\Sigma)$ is bounded away from infinity, $\Lambda_{\min}(\Sigma)$ is bounded away from zero and $\rho(\Phi)$ is bounded away from 1.

\subsection{Proof of  Corollary 1}
Note that $R \geq \sum_{i=1}^q|\bbeta^*_i|^r \geq \sum_{i\in J_{\eta}}|\bbeta^*_i|^r \geq \eta^r|J_{\eta}|$, so we have the upper bound of $|J_{\eta}| $ as
\begin{equation}
|J_{\eta}| \leq R\eta^{-r}.
\label{proof:weak_J}
\end{equation}
Moreover, we upper bound $\|\bbeta^*_{J^C_\eta}\|_1$ using the fact that $\bbeta^*\in \bB(R)$:
\begin{equation}
\|\bbeta^*_{J^C_\eta}\|_1=\sum_{i\in J^C_{\eta}}|\bbeta^*_i|=\sum_{i\in J^C_{\eta}}|\bbeta^*_i|^r|\bbeta^*_i|^{1-r} \leq \sum_{i\in J^C_{\eta}}|\bbeta^*_i|^r\eta^{1-r} \leq R\eta^{1-r}.
\label{proof:weak_beta}
\end{equation}
Substituting these two inequalities into (\ref{proof:weak_l2}), we have
$$\|\bv\|_2 \leq \frac{(w_1+2w_2)}{\alpha} R^{\frac{1}{2}}_r\eta^{\frac{-r}{2}}\lambda_N + 2\sqrt{w_2}\alpha^{-\frac{1}{2}}R^{\frac{1}{2}}\eta^{\frac{1-r}{2}} \lambda_N^{\frac{1}{2}}  + \frac{4w_2\max\{\omega,1\}}{\bQ} R\eta^{1-r} \lambda_N $$
Finally substituting $\eta=\lambda_N/\alpha$ into the above inequality, we have the upper bound of $\|\bv\|_2$ as
$$\|\bv\|_2 \leq (w_1+2w_2+2\sqrt{w_2})\alpha^{\frac{r-2}{2}}R^{\frac{1}{2}}\lambda_N^{\frac{2-r}{2}} + \frac{4w_2\max\{\omega,1\}}{\bQ}\alpha^{r-1}R\lambda_N^{{2-r}}.$$

 \subsection{Proof of Remark (a) of Corollary 1}
		In order to prove $l_r$ ball constraint (\ref{eq:lrball_consist}) is more strict than our weak sparsity constraint (\ref{weaksp_def}), we first prove (\ref{eq:lrball_consist}) implies the (\ref{weaksp_def}), then we show a case in which  (\ref{weaksp_def}) holds but (\ref{eq:lrball_consist}) does not. 
		
		By the choice of $\eta=\lambda_N/\alpha= 4w_1^{-1}\alpha^{-1}\bQ\sqrt{(\log p + 2\log m)/N}$ and inequality (\ref{proof:weak_J}), we have:
		$$|J_{\eta}|\leq 4^{-r}w_1^r\alpha^r\bQ^{-r}\lsk\frac{N}{\log p + 2\log m}\rsk^{r/2}R$$.
		The first equation of $l_r$ ball constraint (\ref{eq:lrball_consist}) implies $R=o\lsk \alpha^{2-r}\bQ^{r-2} \lsk N/(\log p+2\log m)\rsk^{\frac{2-r}{2}}\rsk$, Thus we have:
		$$|J_{\eta}|=o\lsk\lsk\frac{\alpha}{\bQ}\rsk^2\frac{N}{\log p+2\log m}\rsk,$$
		which is the first condition in our weak sparsity constraint (\ref{weaksp_def}). On the other hand, inequality (\ref{proof:weak_beta}) implies
		\begin{equation}
		\|\bbeta^*_{J^C_\eta}\|_1\leq 4^{1-r}w_1^{r-1}\alpha^{r-1}\bQ^{1-r}\lsk\frac{N}{\log p + 2\log m}\rsk^{\frac{r-1}{2}}R.
		\label{proof:coro}
		\end{equation}
		Combing this inequality with the first condition in the $l_r$ ball constraint(\ref{eq:lrball_consist}), we have:
		$$\|\bbeta^*_{J^C_\eta}\|_1=o\lsk \lsk \frac{\alpha}{\bQ} \rsk \sqrt{\frac{N}{\log p+2\log m}}\rsk.$$
		Similarly, combining inequality (\ref{proof:coro}) with the second condition of (\ref{eq:lrball_consist}) will give us:
		$$\|\bbeta^*_{J^C_\eta}\|_1=o\lsk \min\lbk1,\frac{1}{\omega}\rbk \sqrt{\frac{N}{\log p+2\log m}}\rsk.$$
		Thus the second condition of our weak sparsity constraint (\ref{weaksp_def}) holds.
		
		Now we provide a scenario that the $l_r$ ball constraint (\ref{eq:lrball_consist}) does not hold but our weak sparsity constraint (\ref{weaksp_def}) does. Consider a setting in which $\alpha$, $\omega$ and $\bQ$ are bounded from zero and infinity, $m=N$ and $p=1$, then the $l_r$ ball constraint becomes:
		\begin{equation}
		R=o\lsk \lsk\frac{N}{\log N}\rsk^{\frac{2-r}{2}} \rsk,
		\label{proof:lrball_consist2}
		\end{equation}
		while the weak sparsity constraint (\ref{weaksp_def}) becomes:
		\begin{equation}
		|J_{\eta}|=o\lsk\frac{N}{\log N}\rsk\quad \text{and}\quad \|\bbetaJC\|_1=o\lsk\sqrt{\frac{N}{\log N}}\rsk.
		\label{proof:weaksp_def2}
		\end{equation}
		For any give $r\in(0,1]$, we can set $\bbeta^*$ as the following:
		$$\bbeta^*=(\underbrace{1,\dots,1}_{k},\underbrace{1/(N^2\log N),\dots,1/(N^2\log N)}_{N^2-k})',$$
		where $k=\lsk N/(\log N)\rsk^{\frac{2-r}{2}}$. If we set $\eta=0.5$, clearly with large $N$ we have:
		$$|J_{\eta}|=k=o\lsk\frac{N}{\log N}\rsk \;\text{and} \;\|\bbetaJC\|_1=\frac{N^2-k}{N^2\log N}\leq \frac{1}{\log N}=o\lsk\sqrt{\frac{N}{\log N}}\rsk$$
		Thus our weak sparsity constraint (\ref{proof:weaksp_def2}) holds. However, as for the $l_r$ ball constraint (\ref{proof:lrball_consist2}), we have:
		$$=\sum_{i=1}^{N^2}|\beta_i^*|^r\leq \sum_{\beta_i^*=1}^{N^2}|\beta^*|^r=k=\lsk N/(\log N)\rsk^{\frac{2-r}{2}},$$
		in which (\ref{proof:lrball_consist2}) does not hold. If $r=0$, then we can set the true parameter vector $\bbeta^*$ as the follows:
		$$\bbeta^*=(\underbrace{1,\dots,1}_{k},\underbrace{1/(N^2\log N),\dots,1/(N^2\log N)}_{N^2-k})',$$
		where $k=\lsk N/(\log N)\rsk^{\frac{1}{2}}$. Clearly, (\ref{proof:weaksp_def2}) holds but (\ref{proof:lrball_consist2}) does not since $R=\sum_{i=1}^{N^2}I(\beta_i^*\neq0)=N^2$.

\section{Simulation Setting of VAR(2) and VAR(3)}
\setcounter{equation}{0}
\label{subsec:simuVAR23}

To generate the VAR(2) process, we randomly selected 100 sites from the 441 vertices created at the beginning. The magnitude of $\Phi^*_{l,ss'}$ was generated as follows, and its sign was randomly assigned to be -1 or 1 with equal probability. Here we also set $\Sigma=0.01I$. 
\begin{itemize}
	\item[(a)] Exactly sparse:  $|\tilde{\Phi}_{1,ss'}|$ was generated by i.i.d unif(0.1,0.6), $|\tilde{\Phi}_{2,ss'}|$ was generated by i.i.d unif(0.1,0.4). Then set  $|\Phi^*_{1,ss'}|=|\tilde{\Phi}_{1,ss'}|I(d_{ss'}\leq 0.06)$ and $|\Phi^*_{2,ss'}|=|\tilde{\Phi}_{2,ss'}|I(d_{ss'}\leq 0.04)$.
	\item[(b)] Weakly sparse (decay fast): $|\Phi^*_{1,ss'}|=\frac{0.5}{\exp(20\,d_{ss'})}$, $|\Phi^*_{2,ss'}|=\frac{0.3}{\exp(80\,d_{ss'})}$.
	\item[(c)] Weakly sparse (decay slow): $|\Phi^*_{1,ss'}|=\frac{0.3}{\exp(5\,d_{ss'})}$, $|\Phi^*_{2,ss'}|=\frac{0.15}{\exp(20\,d_{ss'})}$.
\end{itemize}
For the simulation of VAR(3) process, we randomly selected 60 sites from the 441 vertices, and used the following three scenarios to generate the magnitudes of the entries in $\Phi_1$, $\Phi_2$ and $\Phi_3$. 
\begin{itemize}
	\item[(a)] Exactly sparse:  $|\tilde{\Phi}_{l,ss'}|$ was generated by i.i.d unif$(0.15,\,0.6-0.1l)$, then set  $|\Phi^*_{l,ss'}|=|\tilde{\Phi}_{l,ss'}|I(d_{ss'}\leq 0.07-0.01l)$.
	\item[(b)] Weakly sparse (decay fast): $|\Phi^*_{l,ss'}|=\frac{0.3}{\exp(25\,l\,d_{ss'})}$.
	\item[(c)] Weakly sparse (decay slow): $|\Phi^*_{l,ss'}|=\frac{0.25}{\exp(10\,l\,d_{ss'})}$.
\end{itemize}

\section{Figures and Tables}

\begin{figure}[H]
	\centering
	\includegraphics[width=0.9\textwidth]{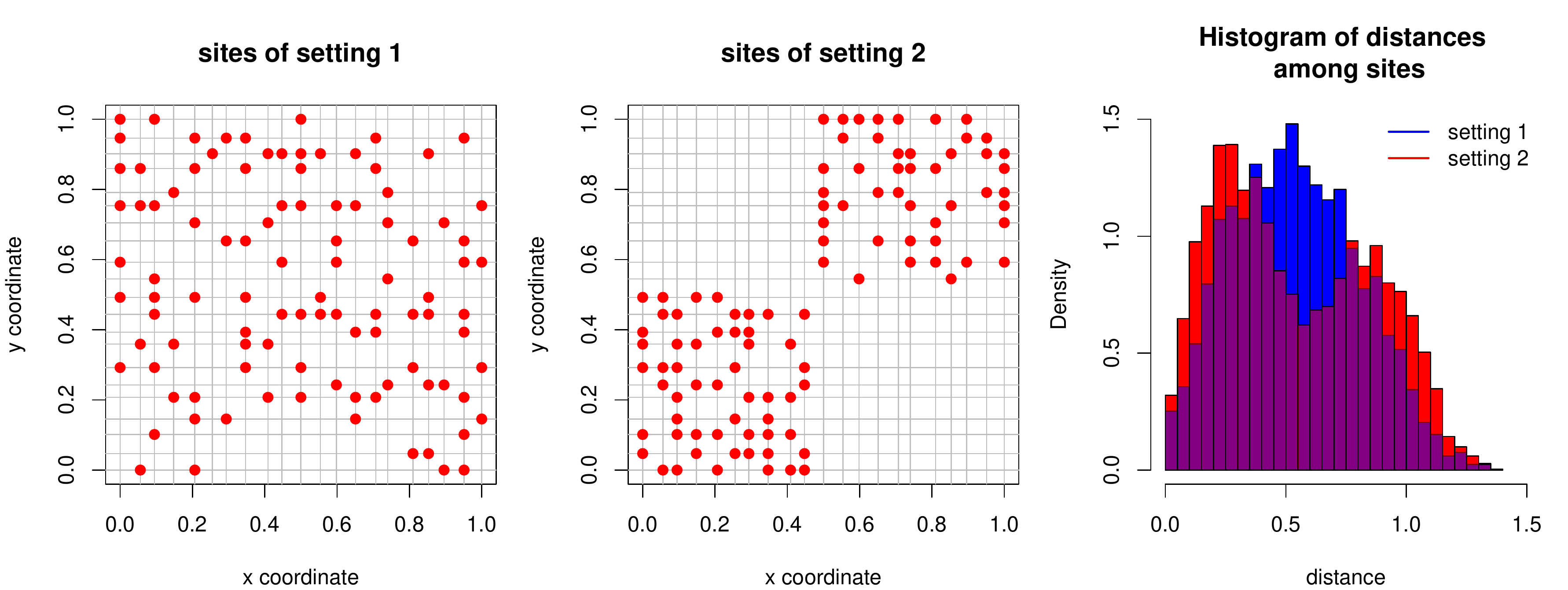}
	\caption{Generated sites in the simulation of VAR(1) and the corresponding histograms of distances among sites. The left two panels plot the locations of the generated sites (red points). The right panel plots the histograms of distance among sites.}
	\label{fig:simu_sites}
\end{figure}

\begin{figure}[H]
	\centering
	\begin{tabular}{@{}ccc@{}}
		\includegraphics[width=.32\textwidth]{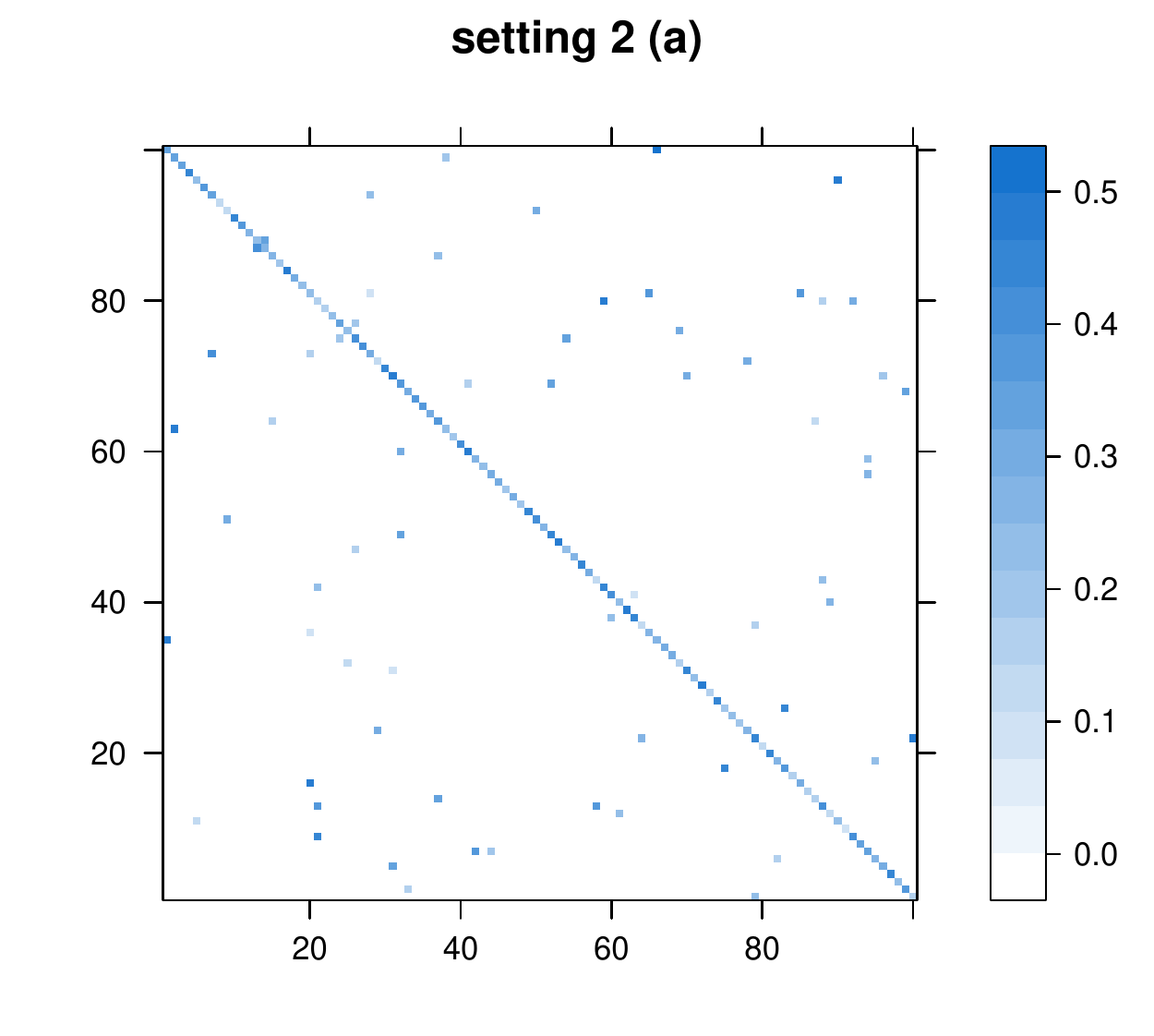} & 
		\includegraphics[width=.32\textwidth]{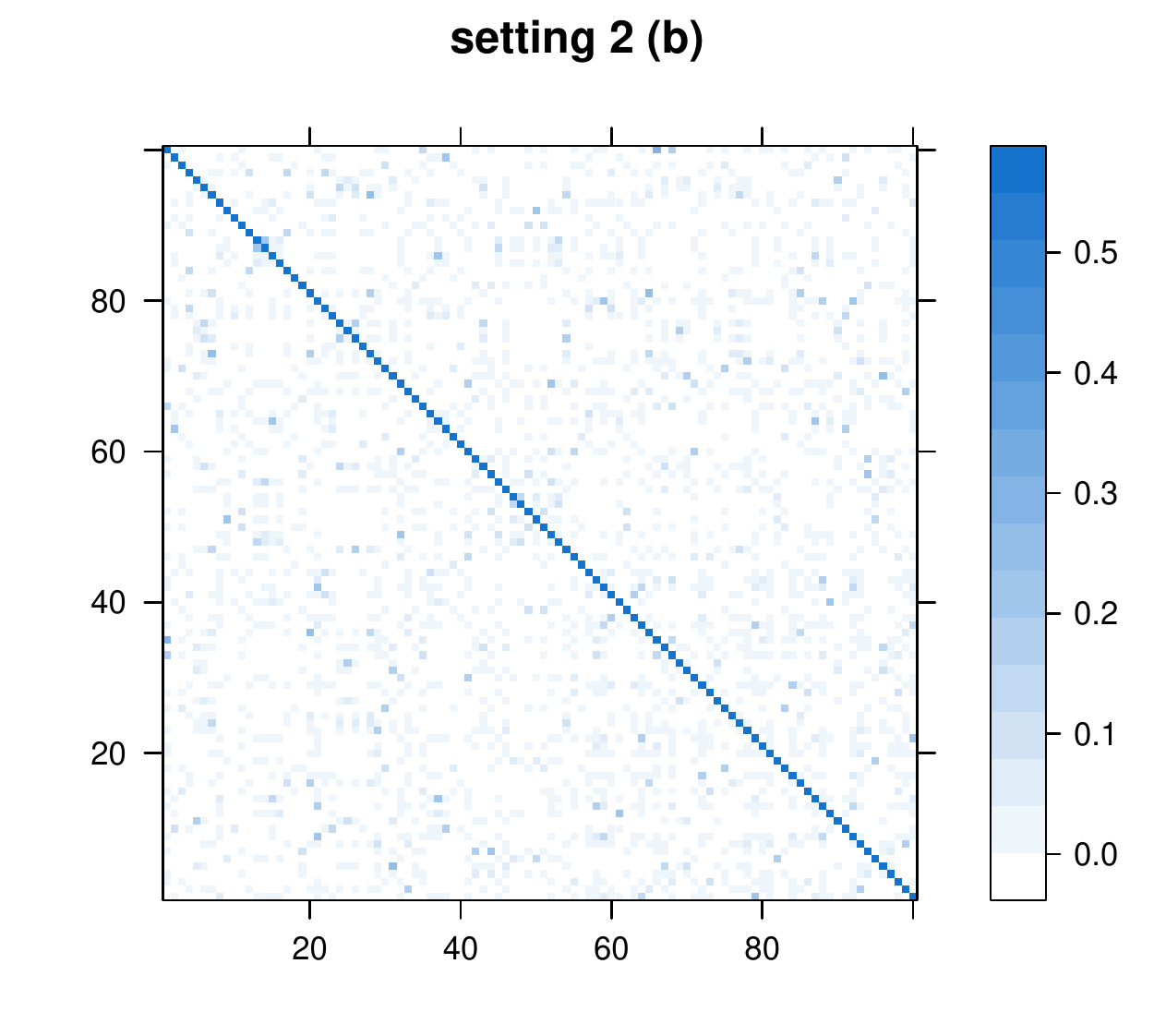} &
		\includegraphics[width=.32\textwidth]{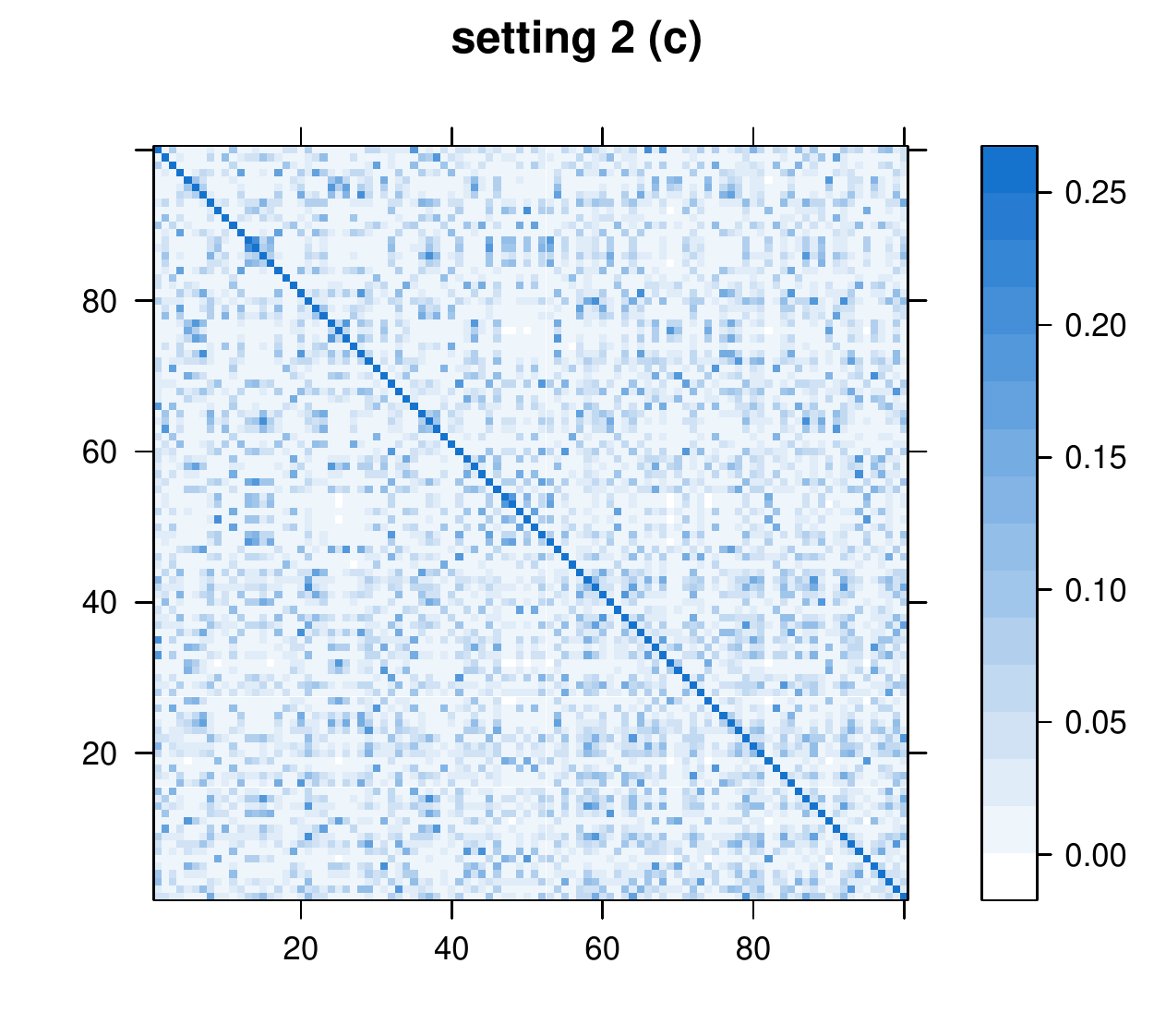}   \\
		\includegraphics[width=.32\textwidth]{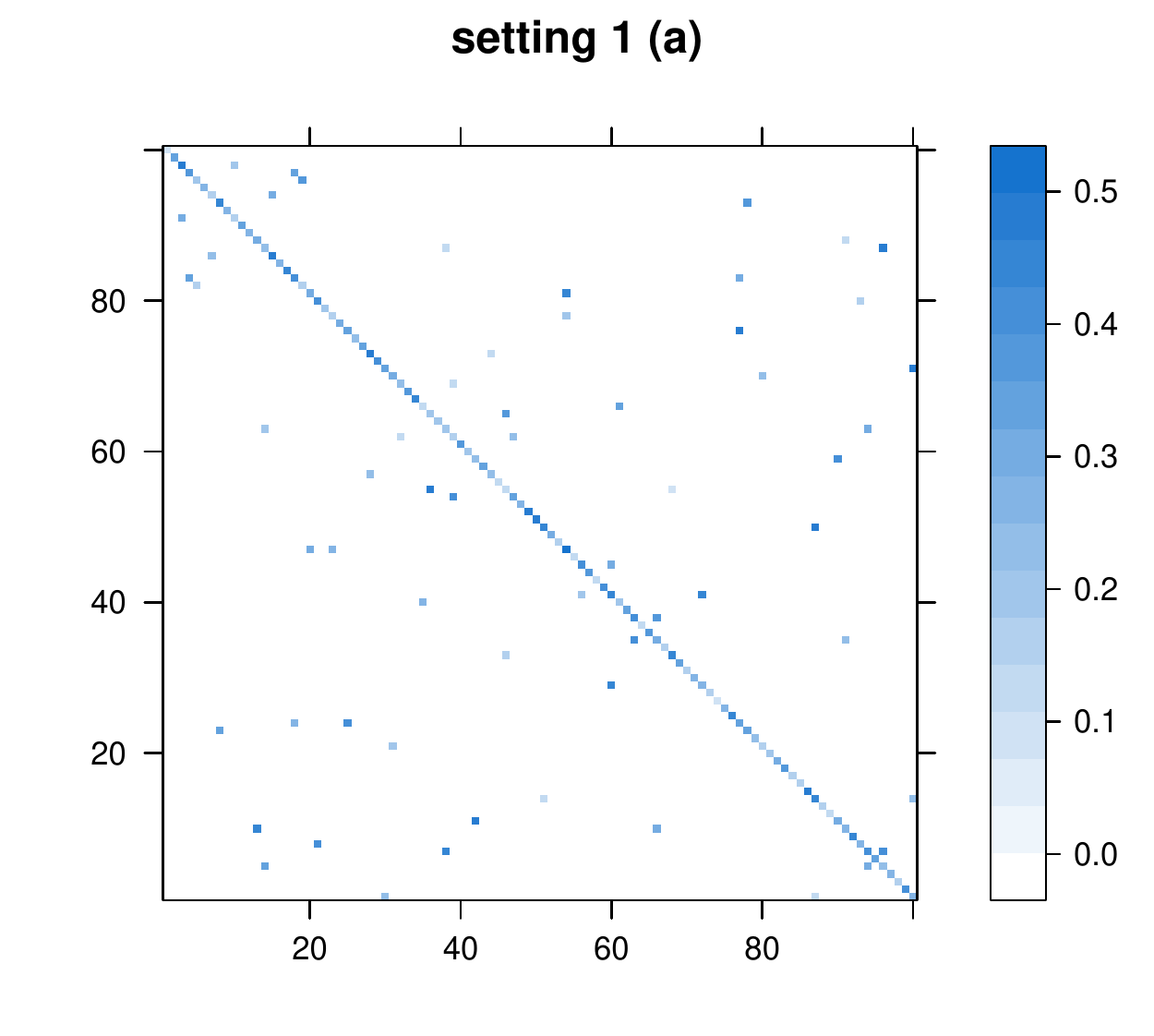} & 
		\includegraphics[width=.32\textwidth]{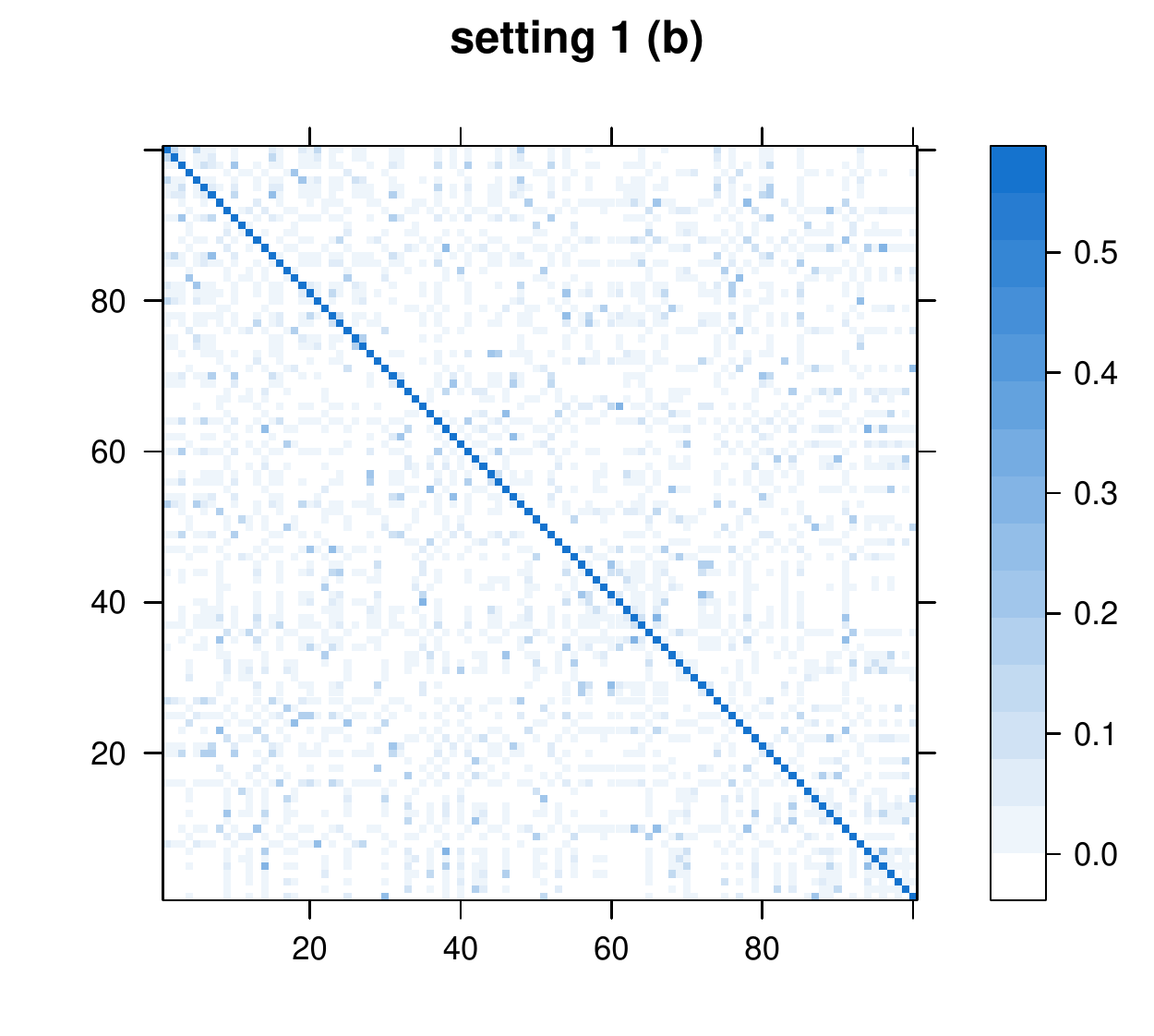} &
		\includegraphics[width=.32\textwidth]{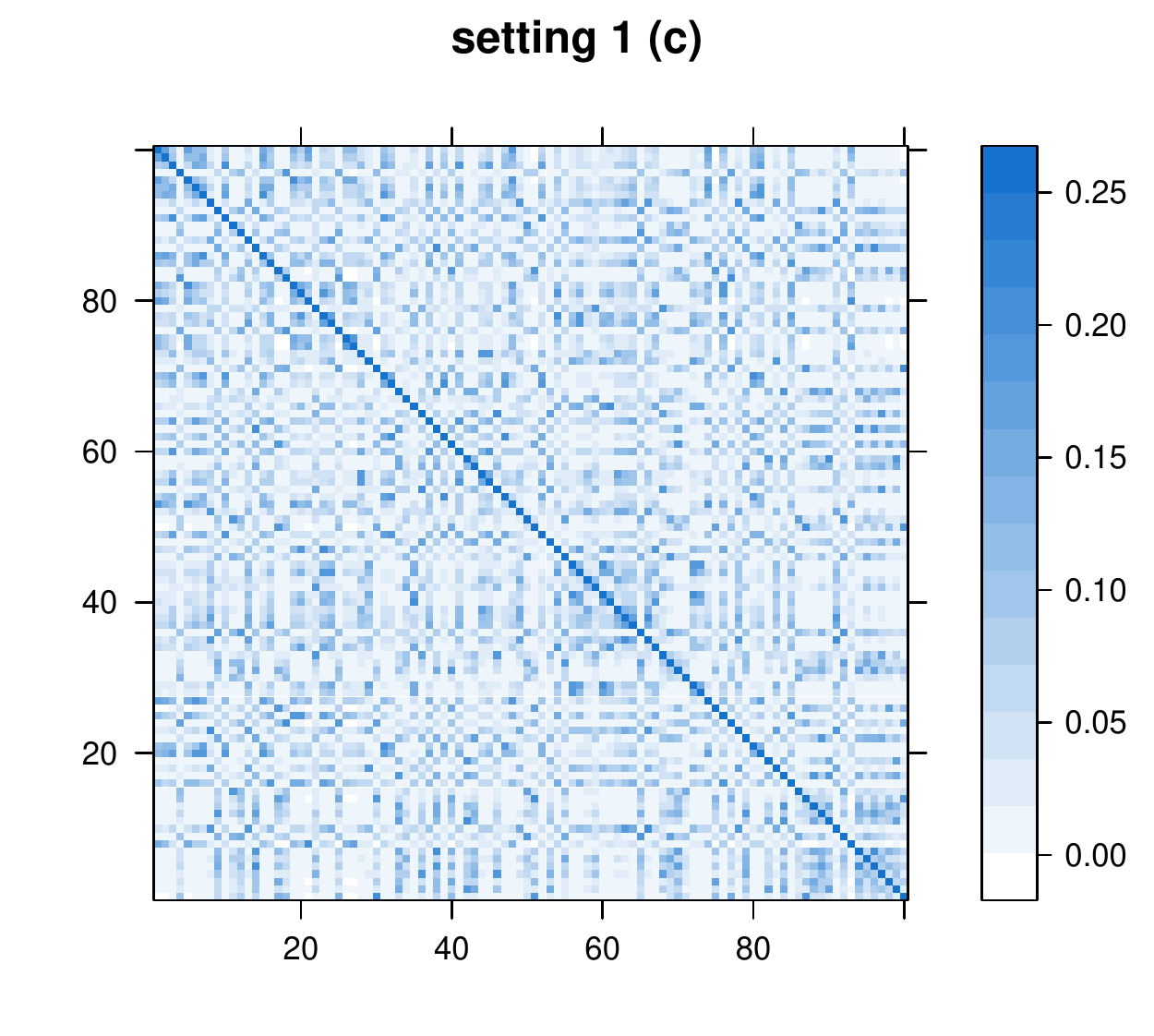}   \\
	\end{tabular}
	\caption{Heat map of $|\Phi_{ss'}|$'s in the simulation of VAR(1) under two setting and  three  scenarios.}
	\label{fig:phi_set1}
\end{figure}

\begin{table}[H]
	\centering
	\caption{Ratio of model fitting criteria of the proposed method, SCAD, MCP and spaLASSO over those of LASSO in setting 1 under three scenarios: (a) exactly sparse scenario, (b) weakly sparse scenario with fast decay, (c) weakly sparse scenario with slow decay. Value below one means the corresponding method outperforms LASSO.}
	\scalebox{0.8}{
		\begin{tabular}{|cccccccc|}
			\hline
			&& \multicolumn{2}{l}{scenario (a)} & \multicolumn{2}{l}{scenario (b)} & \multicolumn{2}{l|}{scenario (c)} \\\cmidrule(r){3-4}\cmidrule(r){5-6}\cmidrule(r){7-8}
			&& mean & se & mean & se & mean & se  \\
			\hline
   \multirow{4}{*}{$l_1$ error ratio} & WLASSO1 & 0.359 & 0.003 & 0.589 & 0.003 & 0.885 & 0.002 \\ 
   & WLASSO2 & 0.359 & 0.003 & 0.589 & 0.003 & 0.885 & 0.002 \\ 
   & SCAD & 1.025 & 0.002 & 0.968 & 0.002 & 1.005 & 0.000 \\ 
   & MCP & 1.028 & 0.003 & 0.936 & 0.002 & 1.011 & 0.001 \\ 
   & spaLASSO & 1.069 & 0.004 & 0.968 & 0.002 & 1.016 & 0.001 \\ 
   \hline
  \multirow{4}{*}{$l_2$ error ratio}  & WLASSO1 & 0.422 & 0.003 & 0.48 & 0.003 & 0.831 & 0.002 \\ 
   & WLASSO2 & 0.423 & 0.003 & 0.48 & 0.003 & 0.831 & 0.002 \\ 
   & SCAD & 1.025 & 0.003 & 0.966 & 0.004 & 1.008 & 0.001 \\ 
   & MCP & 1.049 & 0.004 & 0.965 & 0.004 & 1.018 & 0.001 \\ 
   & spaLASSO & 1.107 & 0.005 & 1.111 & 0.005 & 1.026 & 0.001 \\ 
   \hline
   \multirow{4}{*}{PFZ ratio} & WLASSO1 & 0.049 & 0.003 & -- & -- & -- & -- \\ 
   & WLASSO2 & 0.05 & 0.003 & -- & -- & -- & -- \\ 
   & SCAD & 1.038 & 0.009 & -- & -- & -- & -- \\ 
   & MCP & 1.221 & 0.01 & -- & -- & -- & -- \\ 
   & spaLASSO & 1.478 & 0.013 & -- & -- & -- & -- \\ 
   \hline
   \multirow{4}{*}{PFNZ ratio} & WLASSO1 & 0.245 & 0.069 & -- & -- & -- & -- \\ 
   & WLASSO2 & 0.251 & 0.069 & -- & -- & -- & -- \\ 
   & SCAD & 0.245 & 0.014 & -- & -- & -- & -- \\ 
   & MCP & 0.058 & 0.006 & -- & -- & -- & -- \\ 
   & spaLASSO & 0.014 & 0.001 & -- & -- & -- & -- \\ 
   \hline
		\end{tabular}
	}
	\label{table:simu1}
\end{table}

\begin{figure}[H]
	\centering
	\begin{tabular}{@{}c@{}}
		\includegraphics[width=1\textwidth]{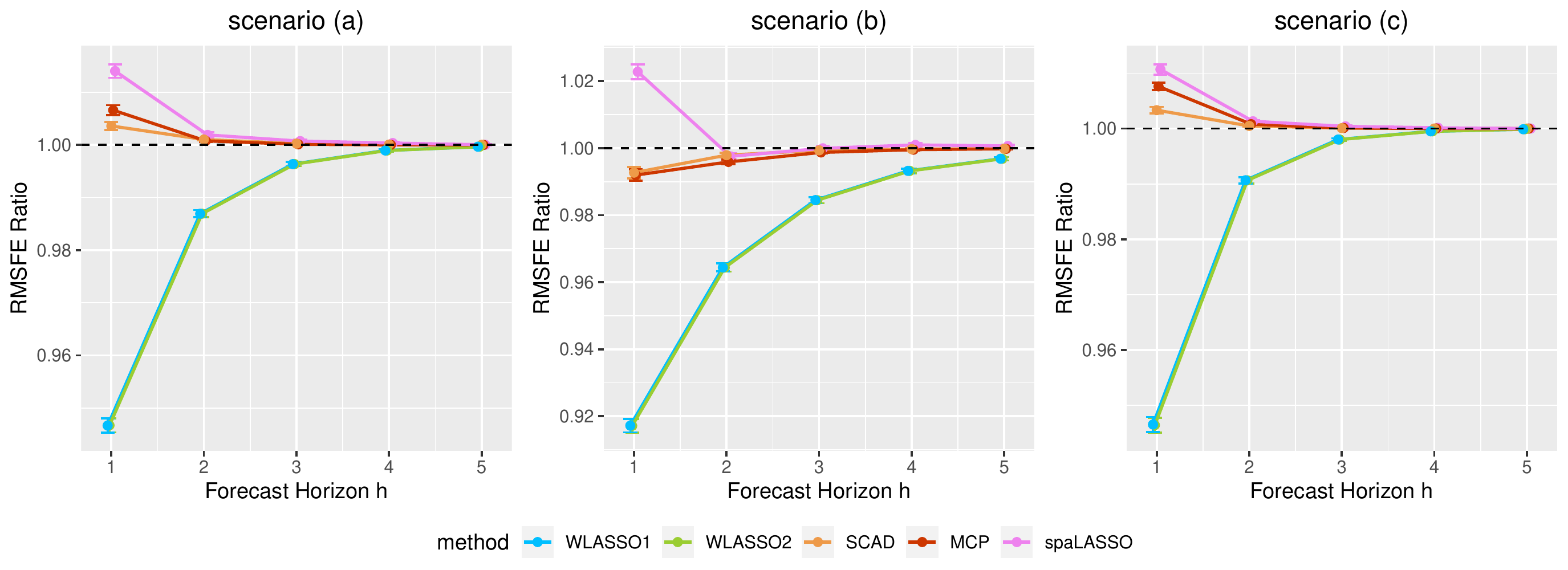}    
	\end{tabular}	
	\caption{Ratio of RMSFE of the proposed method, SCAD, MCP and spaLASSO over that of LASSO in setting 1 under three scenarios: (a) exactly sparse scenario, (b) weakly sparse scenario with fast decay, (c) weakly sparse scenario with slow decay. The solid line plots the means of RMSFE ratios in the 100 replicates with error bars standing for twice its standard errors. The dashed horizontal line is at ratio one, and values below it indicate the corresponding method has smaller RMSFE than LASSO.}
	\label{fig:forecast1}
\end{figure}

\begin{table}[H]
	\centering
	\caption{Ratio of model fitting criteria of different methods over those of LASSO in setting 2 under three scenarios: (a) exactly sparse scenario, (b) weakly sparse scenario with fast decay, (c) weakly sparse scenario with slow decay. Value below one means the proposed method outperforms LASSO.}
	\scalebox{0.8}{
		\begin{tabular}{|cccccccc|}
			\hline
			&& \multicolumn{2}{l}{scenario (a)} & \multicolumn{2}{l}{scenario (b)} & \multicolumn{2}{l|}{scenario (c)} \\\cmidrule(r){3-4}\cmidrule(r){5-6}\cmidrule(r){7-8}
			&& mean & se & mean & se & mean & se  \\
			\hline
			 \multirow{4}{*}{$l_1$ error ratio} & WLASSO1 & 0.368 & 0.003 & 0.636 & 0.002 & 0.861 & 0.001 \\ 
			 & WLASSO2 & 0.368 & 0.003 & 0.635 & 0.002 & 0.861 & 0.002 \\ 
			 & SCAD & 1.031 & 0.002 & 0.982 & 0.002 & 1.015 & 0.001 \\ 
			 & MCP & 1.03 & 0.004 & 0.969 & 0.002 & 1.032 & 0.001 \\ 
			 & spaLASSO & 1.079 & 0.004 & 1.015 & 0.002 & 1.043 & 0.001 \\ 
			 \hline
			 \multirow{4}{*}{$l_2$ error ratio} & WLASSO1 & 0.433 & 0.003 & 0.54 & 0.002 & 0.833 & 0.002 \\ 
			 & WLASSO2 & 0.433 & 0.003 & 0.54 & 0.002 & 0.833 & 0.002 \\ 
			 & SCAD & 1.033 & 0.003 & 0.985 & 0.004 & 1.019 & 0.001 \\ 
			 & MCP & 1.059 & 0.004 & 0.994 & 0.003 & 1.04 & 0.001 \\ 
			 & spaLASSO & 1.128 & 0.004 & 1.134 & 0.004 & 1.056 & 0.002 \\ 
			 \hline
			  \multirow{4}{*}{PFZ ratio} & WLASSO1 & 0.065 & 0.003 & -- & -- & -- & -- \\ 
			 & WLASSO2 & 0.065 & 0.003 & -- & -- & -- & -- \\ 
			 & SCAD & 1.04 & 0.008 & -- & -- & -- & -- \\ 
			 & MCP & 1.233 & 0.01 & -- & -- & -- & -- \\ 
			 & spaLASSO & 1.519 & 0.014 & -- & -- & -- & -- \\ 
			 \hline
			  \multirow{4}{*}{PFNZ ratio} & WLASSO1 & 0.148 & 0.022 & -- & -- & -- & -- \\ 
			 & WLASSO2 & 0.154 & 0.023 & -- & -- & -- & -- \\ 
			 & SCAD & 0.306 & 0.018 & -- & -- & -- & -- \\ 
			 & MCP & 0.067 & 0.006 & -- & -- & -- & -- \\ 
			 & spaLASSO & 0.015 & 0.001 & -- & -- & -- & -- \\ 
			\hline
		\end{tabular}
	}
	\label{table:simu2}
\end{table}

\begin{figure}[H]
	\centering
		\includegraphics[width=1\textwidth]{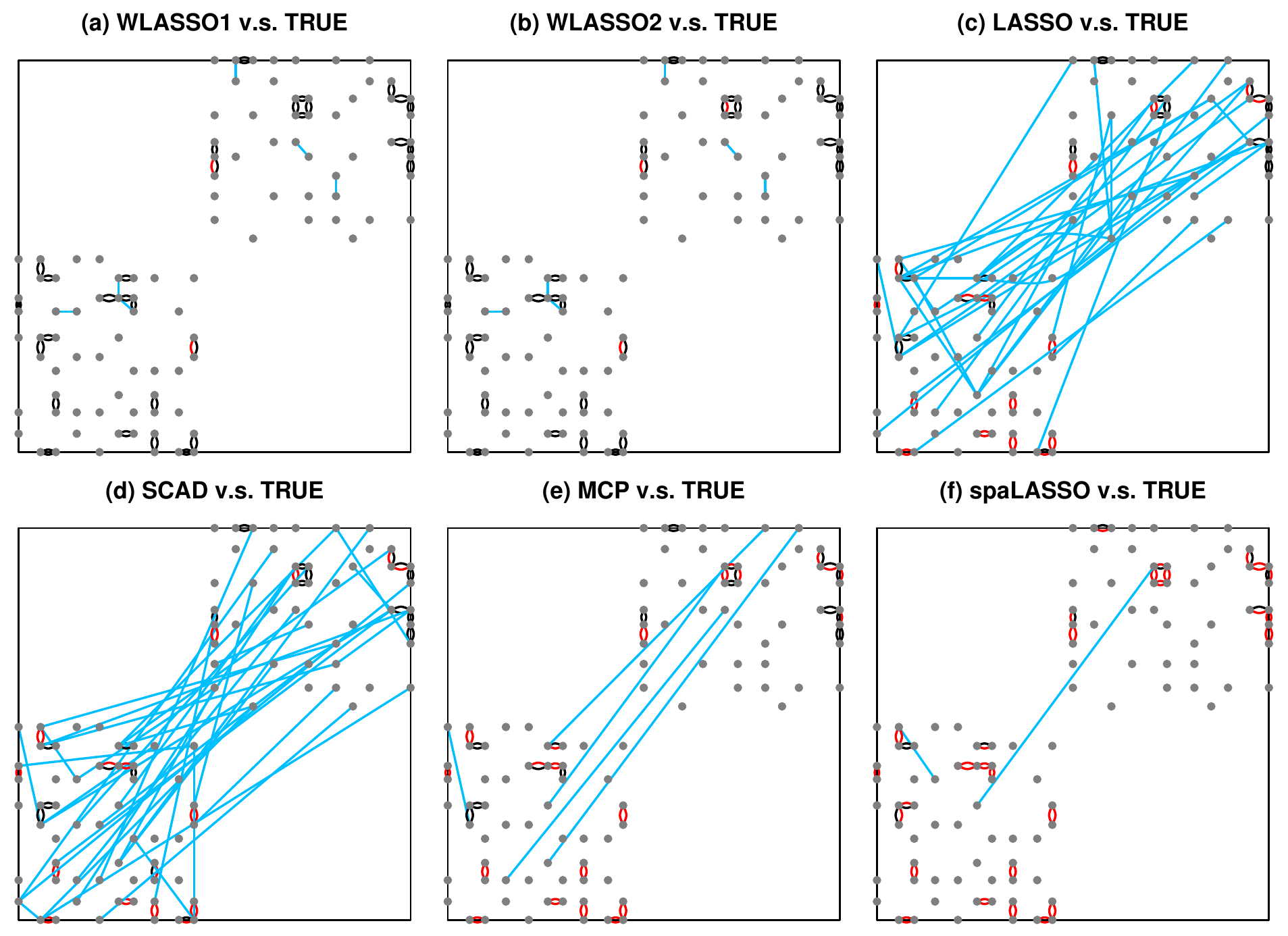}
	\caption{Network detection from different methods of one randomly selected replicate in exactly sparse scenario (a) of setting 2. In detail, if both $\Phi^*_{ss'}$ and its estimator $\hat{\Phi}_{ss'}$ are nonzero, a black edge is drawn to connect site $i$ and site $j$. If $\Phi^*_{ss'}$ is not zero but $\hat{\Phi}_{ss'}$ is zero, the edge is red. If $\Phi^*_{ss'}$ is zero but $\hat{\Phi}_{ss'}$ is not zero, the edge is blue.} 
	\label{fig:net_setting2}
\end{figure}

\begin{figure}[H]
	\centering
	\begin{tabular}{@{}c@{}}
		\includegraphics[width=1\textwidth]{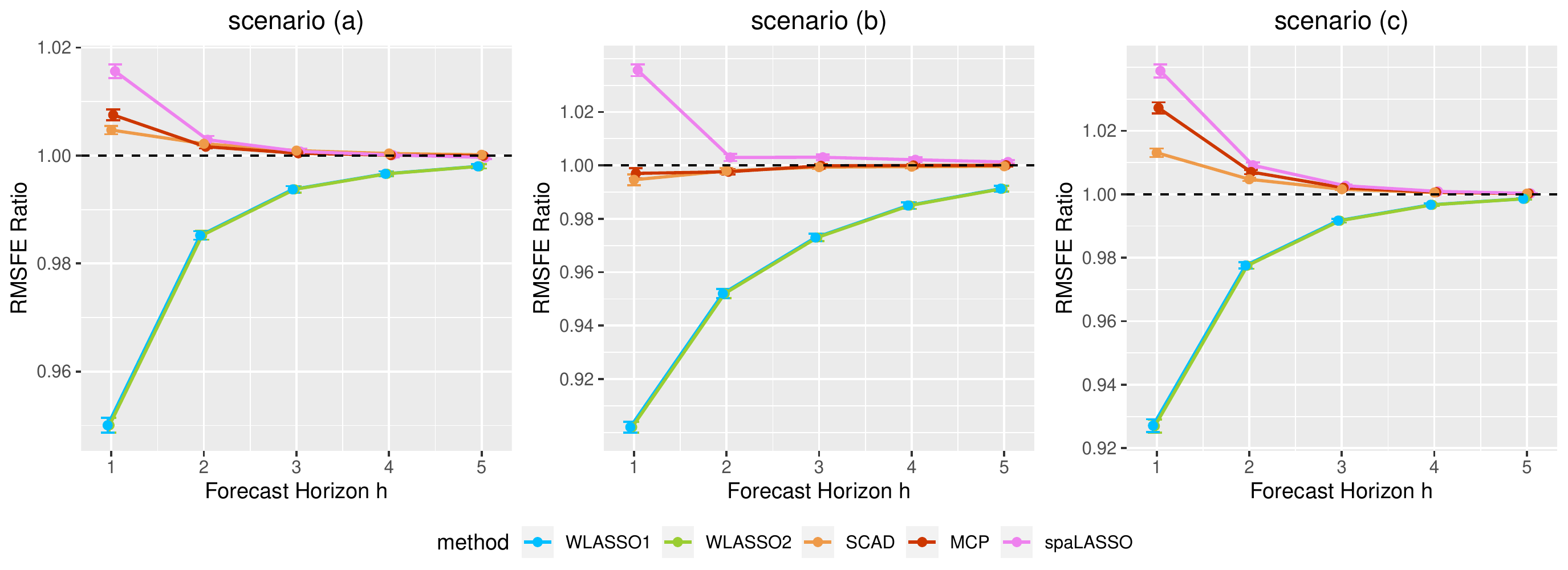}    
	\end{tabular}	
	\caption{Ratio of RMSFE of different methods over that of LASSO for each horizon in setting 2 under three scenarios: (a) exactly sparse scenario, (b) weakly sparse scenario with fast decay, (c) weakly sparse scenario with slow decay. In each panel, the dashed horizontal line is at ratio equaling to one, and the solid line plots the means of RMSFE ratios in the 100 replicates with error bars standing for twice of its standard errors. }
	\label{fig:forecast2}
\end{figure}

\begin{table}[H]
	\centering
	\caption{Ratio of model fitting criteria of different methods over those of LASSO in the simulation of VAR(2) under three scenarios: (a) exactly sparse scenario, (b) weakly sparse scenario with fast decay, (c) weakly sparse scenario with slow decay.}
\scalebox{0.8}{
	\begin{tabular}{|cccccccc|}
		\hline
		&& \multicolumn{2}{l}{scenario (a)} & \multicolumn{2}{l}{scenario (b)} & \multicolumn{2}{l|}{scenario (c)} \\\cmidrule(r){3-4}\cmidrule(r){5-6}\cmidrule(r){7-8}
		&& mean & se & mean & se & mean & se  \\
		\hline
		 \multirow{4}{*}{$l_1$ error ratio} & WLASSO1 & 0.358 & 0.003 & 0.508 & 0.002 & 0.848 & 0.001 \\ 
		 & WLASSO2 & 0.358 & 0.003 & 0.508 & 0.002 & 0.848 & 0.001 \\ 
		 & SCAD & 0.988 & 0.001 & 0.97 & 0.001 & 1.022 & 0.001 \\ 
		 & MCP & 0.974 & 0.002 & 0.966 & 0.002 & 1.04 & 0.001 \\ 
		 & spaLASSO & 1.022 & 0.002 & 1.023 & 0.002 & 1.061 & 0.001 \\ 
		 \hline
		 \multirow{4}{*}{$l_2$ error ratio} & WLASSO1 & 0.428 & 0.004 & 0.448 & 0.002 & 0.799 & 0.001 \\ 
		 & WLASSO2 & 0.428 & 0.004 & 0.448 & 0.002 & 0.798 & 0.001 \\ 
		 & SCAD & 1.005 & 0.002 & 1.001 & 0.003 & 1.031 & 0.001 \\ 
		 & MCP & 1.026 & 0.003 & 1.023 & 0.002 & 1.058 & 0.001 \\ 
		 & spaLASSO & 1.099 & 0.003 & 1.109 & 0.003 & 1.096 & 0.002 \\ 
		 \hline
		  \multirow{4}{*}{PFZ ratio} & WLASSO1 & 0.063 & 0.006 & -- & -- & -- & -- \\ 
		 & WLASSO2 & 0.063 & 0.006 & -- & -- & -- & -- \\ 
		 & SCAD & 0.99 & 0.005 & -- & -- & -- & -- \\ 
		 & MCP & 1.114 & 0.005 & -- & -- & -- & -- \\ 
		 & spaLASSO & 1.322 & 0.007 & -- & -- & -- & -- \\ 
		 \hline
		  \multirow{4}{*}{PFNZ ratio} & WLASSO1 & 0.075 & 0.006 & -- & -- & -- & -- \\ 
		 & WLASSO2 & 0.078 & 0.007 & -- & -- & -- & -- \\ 
		 & SCAD & 0.456 & 0.022 & -- & -- & -- & -- \\ 
		 & MCP & 0.087 & 0.004 & -- & -- & -- & -- \\ 
		 & spaLASSO & 0.01 & 0.001 & -- & -- & -- & -- \\ 
		 \hline
	\end{tabular}
}
	\label{table:simu_fitp2}
\end{table}

\begin{table}[H]
	\centering
	\caption{Ratio of model fitting criteria of different methods over those of LASSO in the simulation of VAR(3) under three scenarios: (a) exactly sparse scenario, (b) weakly sparse scenario with fast decay, (c) weakly sparse scenario with slow decay.} 
	\scalebox{0.8}{
		\begin{tabular}{|cccccccc|}
			\hline
			&& \multicolumn{2}{l}{scenario (a)} & \multicolumn{2}{l}{scenario (b)} & \multicolumn{2}{l|}{scenario (c)} \\\cmidrule(r){3-4}\cmidrule(r){5-6}\cmidrule(r){7-8}
			&& mean & se & mean & se & mean & se  \\
			\hline
			\multirow{4}{*}{$l_1$ error ratio} & WLASSO1 & 0.524 & 0.008 & 0.407 & 0.003 & 0.708 & 0.006 \\ 
			& WLASSO2 & 0.524 & 0.008 & 0.407 & 0.003 & 0.709 & 0.006 \\ 
			& SCAD & 1.013 & 0.002 & 1.012 & 0.004 & 1.002 & 0 \\ 
			& MCP & 1.01 & 0.002 & 1.016 & 0.004 & 1.006 & 0.001 \\ 
			& spaLASSO & 1.028 & 0.003 & 1.008 & 0.005 & 1.02 & 0.001 \\ 
			\hline
			\multirow{4}{*}{$l_2$ error ratio} & WLASSO1 & 0.598 & 0.008 & 0.421 & 0.003 & 0.645 & 0.008 \\ 
			& WLASSO2 & 0.597 & 0.008 & 0.421 & 0.003 & 0.645 & 0.008 \\ 
			& SCAD & 1.022 & 0.002 & 1.024 & 0.004 & 1.004 & 0.001 \\ 
			& MCP & 1.037 & 0.003 & 1.038 & 0.004 & 1.009 & 0.001 \\ 
			& spaLASSO & 1.067 & 0.003 & 1.034 & 0.004 & 1.031 & 0.001 \\ 
			\hline
			\multirow{4}{*}{PFZ ratio} & WLASSO1 & 0.28 & 0.017 & -- & -- & -- & -- \\ 
			& WLASSO2 & 0.28 & 0.016 & -- & -- & -- & -- \\ 
			& SCAD & 1.02 & 0.007 & -- & -- & -- & -- \\ 
			& MCP & 1.081 & 0.008 & -- & -- & -- & -- \\ 
			& spaLASSO & 1.122 & 0.008 & -- & -- & -- & -- \\ 
			\hline
			\multirow{4}{*}{PFZ ratio} & WLASSO1 & 0.388 & 0.067 & -- & -- & -- & -- \\ 
			& WLASSO2 & 0.397 & 0.071 & -- & -- & -- & -- \\ 
			& SCAD & 0.195 & 0.015 & -- & -- & -- & -- \\ 
			& MCP & 0.048 & 0.004 & -- & -- & -- & -- \\ 
			& spaLASSO & 0.007 & 0.001 & -- & -- & -- & -- \\ 
			\hline
		\end{tabular}
	}
	\label{table:simu_fitp3}
\end{table}

\begin{figure}[H]
	\centering
	\includegraphics[width=1\textwidth]{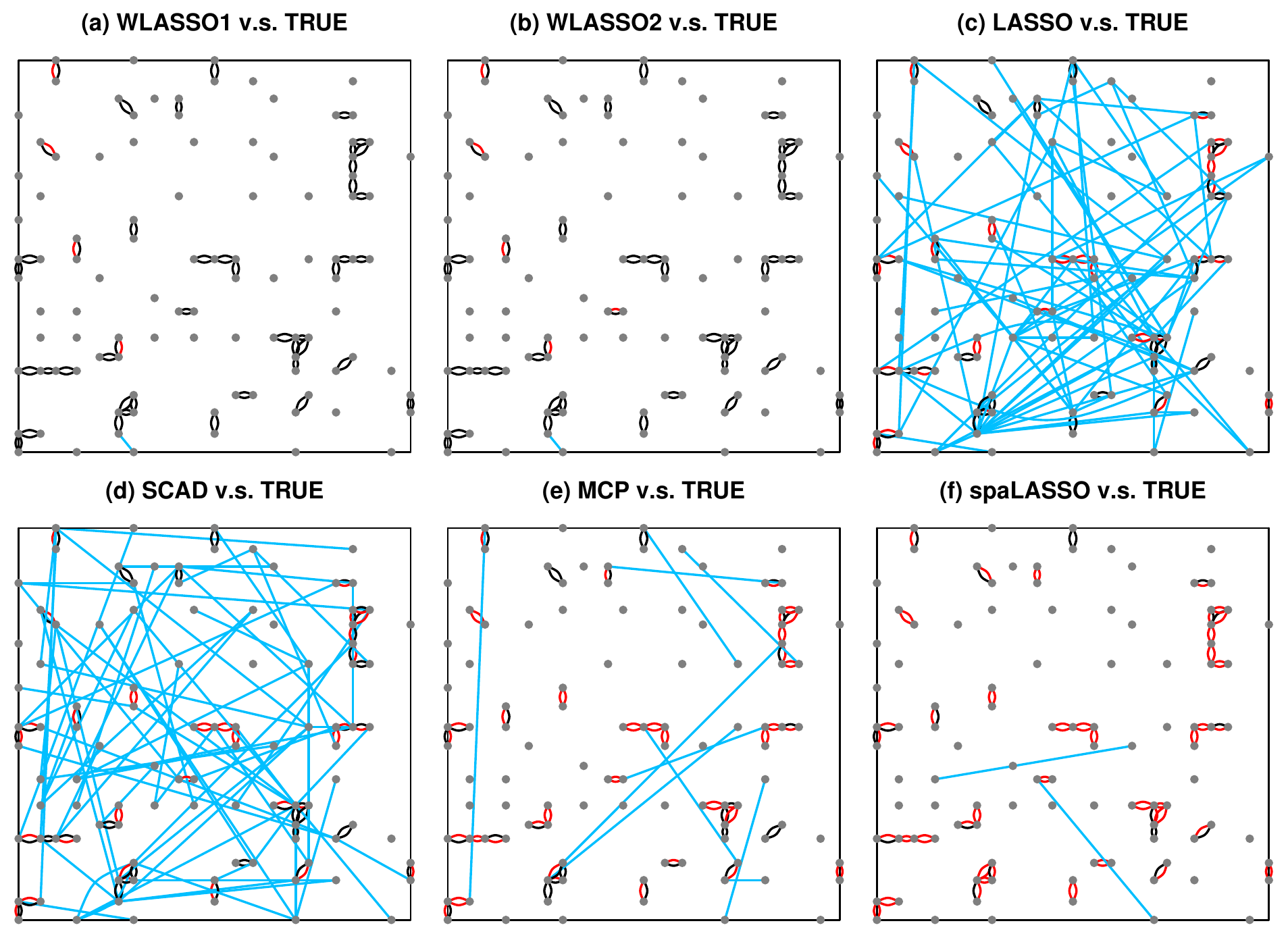}
	\caption{Network detection from different methods of one randomly selected replicate in exactly sparse scenario (a) of VAR(2). In detail, if both $\Phi^*_{ss'}$ and its estimator $\hat{\Phi}_{ss'}$ are nonzero, a black edge is drawn to connect site $i$ and site $j$. If $\Phi^*_{ss'}$ is not zero but $\hat{\Phi}_{ss'}$ is zero, the edge is red. If $\Phi^*_{ss'}$ is zero but $\hat{\Phi}_{ss'}$ is not zero, the edge is blue. }
	\label{fig:net_p2}
\end{figure}

\begin{figure}[H]
	\centering
	\includegraphics[width=1\textwidth]{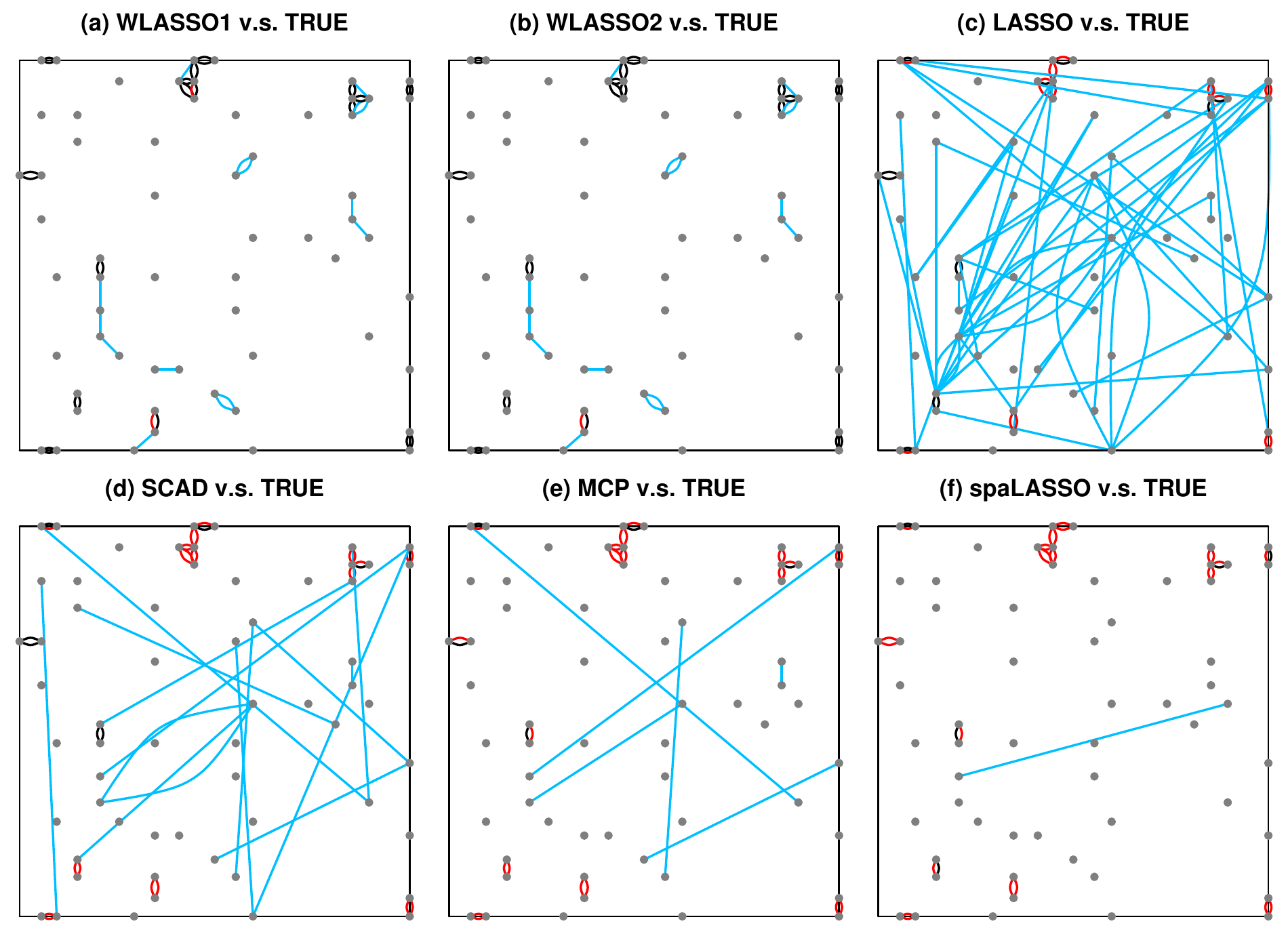}
	\caption{Network detection from different methods of one randomly selected replicate in exactly sparse scenario (a) of VAR(3). In detail, if both $\Phi^*_{ss'}$ and its estimator $\hat{\Phi}_{ss'}$ are nonzero, a black edge is drawn to connect site $i$ and site $j$. If $\Phi^*_{ss'}$ is not zero but $\hat{\Phi}_{ss'}$ is zero, the edge is red. If $\Phi^*_{ss'}$ is zero but $\hat{\Phi}_{ss'}$ is not zero, the edge is blue.} 
	\label{fig:net_p3}
\end{figure}

\begin{figure}[H]
	\centering
	\begin{tabular}{@{}c@{}}
		\includegraphics[width=1\textwidth]{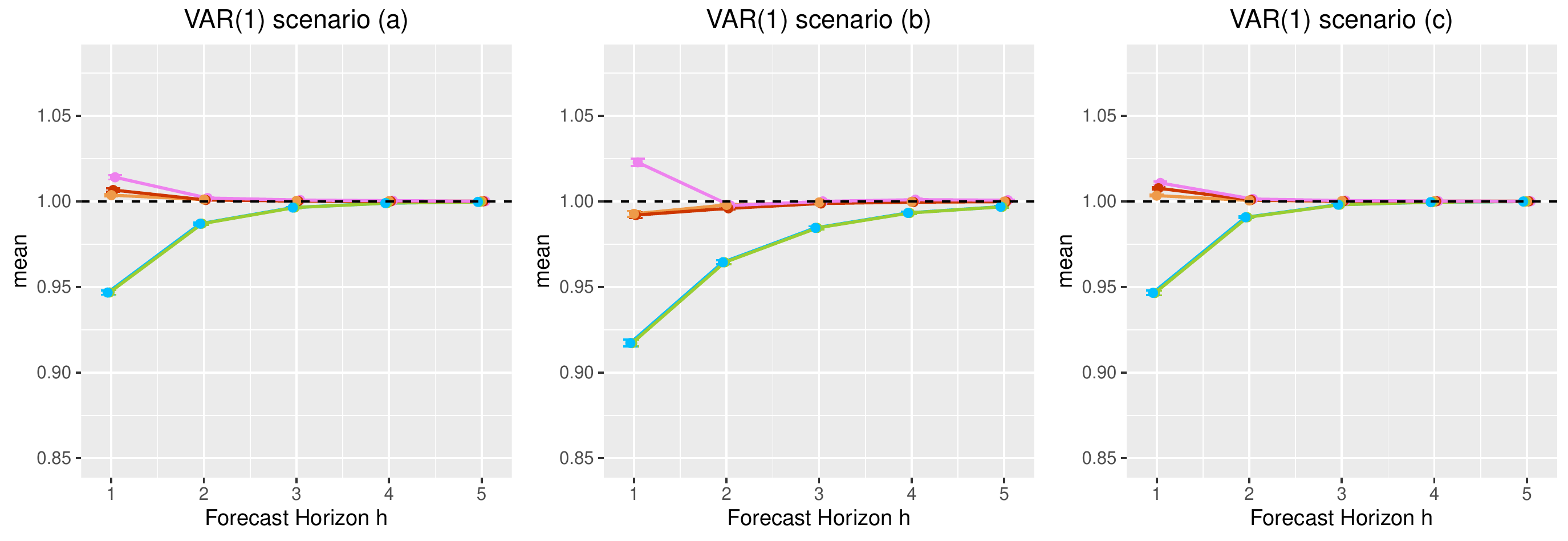}   \\
		\includegraphics[width=1\textwidth]{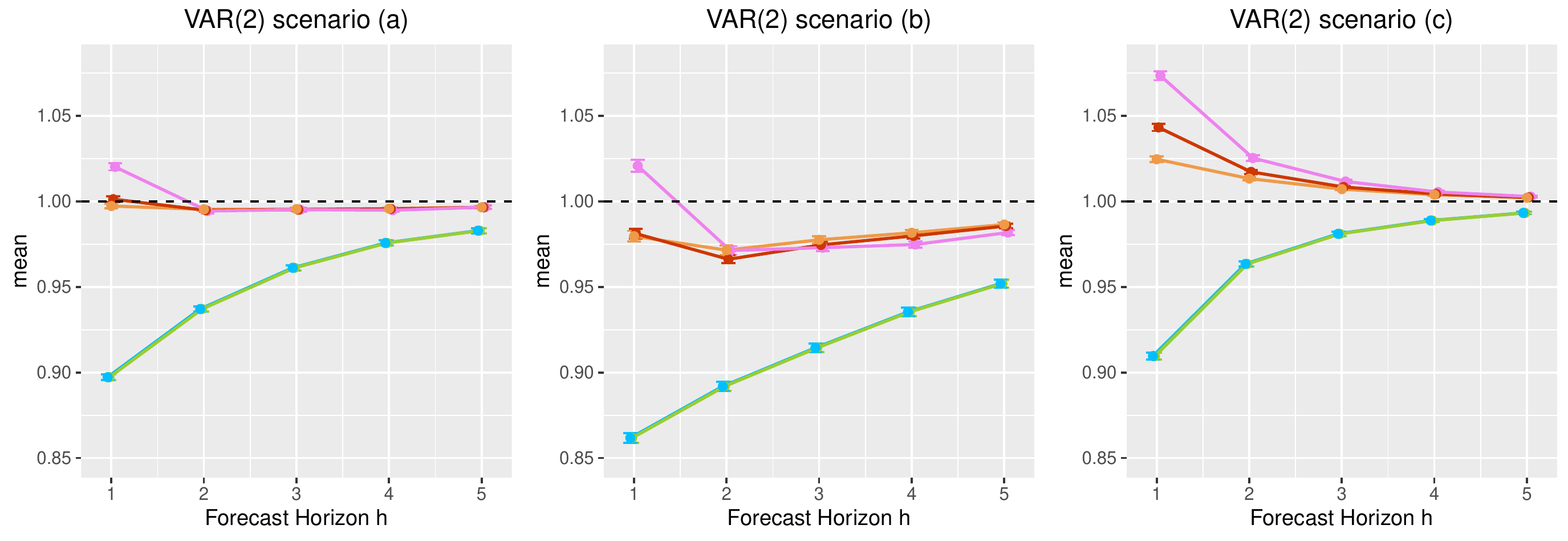}   \\
		\includegraphics[width=1\textwidth]{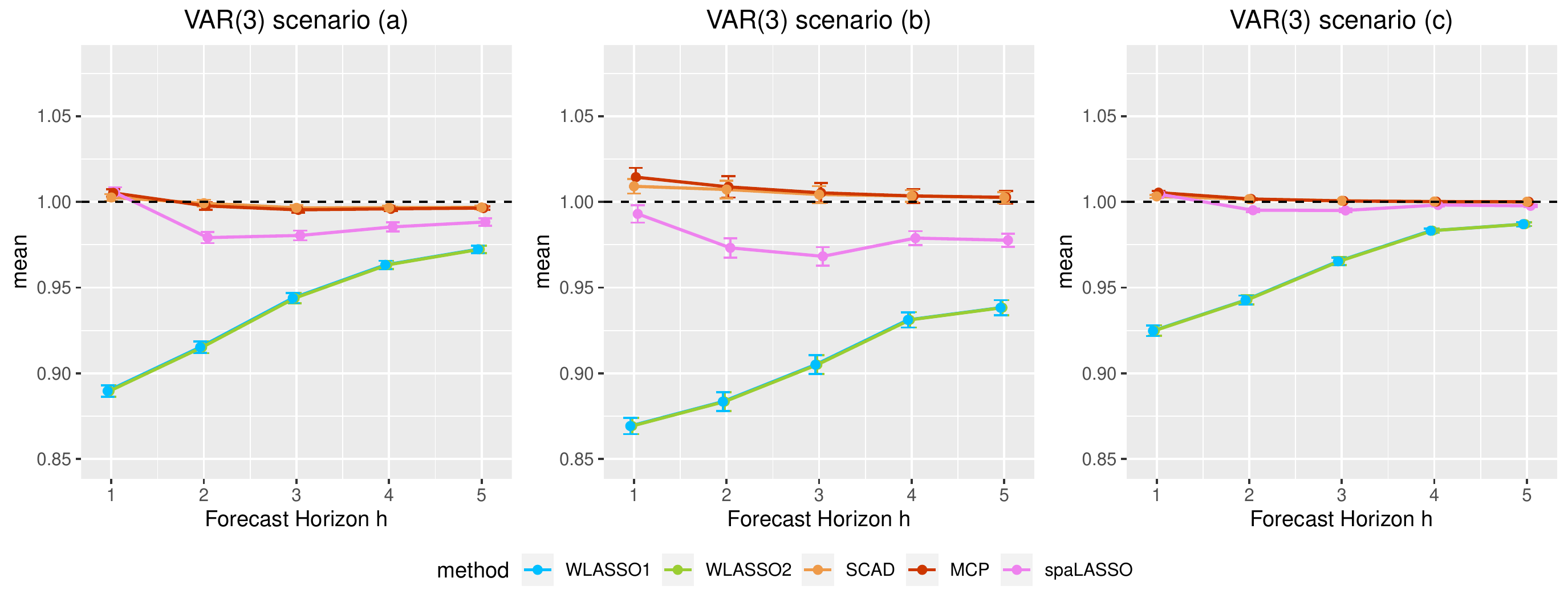}   
	\end{tabular} 
	\caption{Ratio of RMSFE of different methods over that of LASSO for VAR(1), VAR(2) and VAR(3) under three scenarios. In each panel, the dashed horizontal line is at ratio equaling to one, and the solid line plots the means of RMSFE ratios in the 100 replicates with error bars standing for twice of its standard errors.} 
	\label{fig:forecast_p23}
\end{figure}

\begin{figure}[H]
	\centering
	\includegraphics[width=0.9\textwidth]{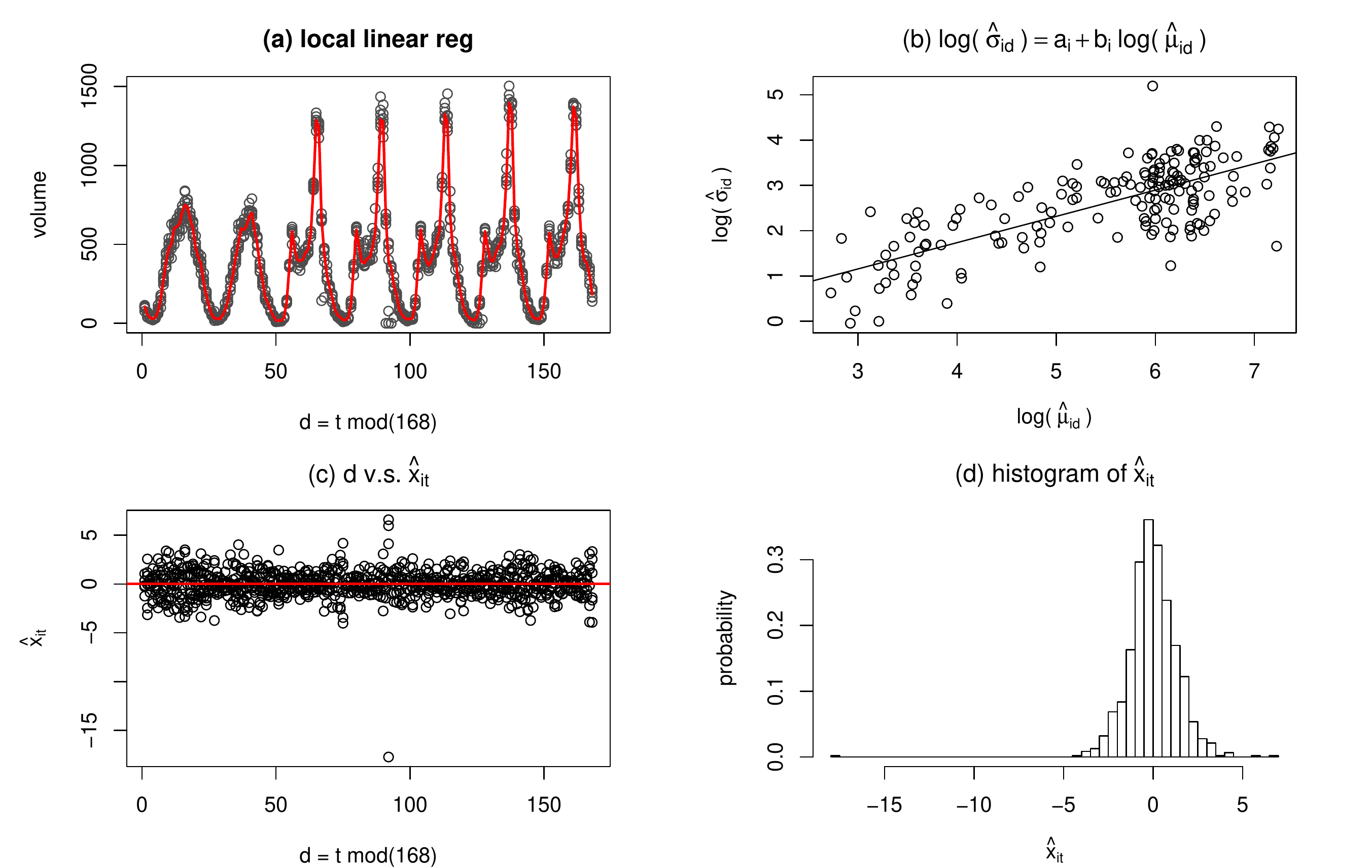}
	\caption{Stage 1 result of site ``IA-5 EAST of SW 9th-EB". Panel (a) is the scatterplot of $d=t\,\text{mod}\,(168)$ versus $z_{it}$ (grey points) and the local linear kernel regression (red curve); Panel (b) plots $\log(\hat{\mu}_{id})$ versus $\log(\hat{\sigma}_{id})$ and the linear regression line; Panel (c) plots $d$ versus estimated $\hat{x}_{it}$; Panel (d) gives the histogram of $\{\hat{x}_{it}\}$.}
	\label{fig:reg}
\end{figure}

\begin{table}[H]
	\centering
	\caption{Number of observations in training, validation and test dataset.}
	\begin{tabular}{|cccccc|}
		\hline
		& \multicolumn{2}{c}{VAR} & \multicolumn{3}{c|}{LASSO and WLASSO} \\ \cmidrule(lr){2-3}\cmidrule(lr){4-6}
		& train & test & train & validation & test   \\ 
		\hline
		weekday peak & 300 & 150 & 150 & 150 & 150 \\
		weekday off-peak & 180 & 90 & 90 & 90 & 90 \\
		weekend peak & 128 & 52 & 76 & 52 & 52 \\
		weekend off-peak & 103 & 44 & 59 & 44 & 44 \\
		\hline
	\end{tabular}
	\label{table:nobs}
\end{table}

\begin{table}[H]
	\centering
	\caption{Selected VAR order $p$ from different method via cross validation}
	\scalebox{0.85}{
			\begin{tabular}{|ccccc|}
			\hline
			& weekday peak & weekday off-peak & weekend peak & weekend off-peak \\ 
			\hline
			LASSO &   1 &    1 &   5 &   2 \\ 
			WLASSO1 &   1 &   2 &   2 &   1 \\ 
			\hline
		\end{tabular}
	}
	\label{table:real_order}
\end{table}

\begin{table}[H]
	\centering
	\caption{$h$-step ahead RMSFE for each method and sub-period. The DM test is for comparison between WLASSO1 and LASSO. Bold value means the gain of WLASSO1 over LASSO is significant under significant level 0.05. }
	\scalebox{0.8}{
		\begin{tabular}{|ccccccccc|}
			\hline
			& \multicolumn{4}{c}{weekday peak time } & \multicolumn{4}{c|}{weekday off-peak time} \\
			\cmidrule(lr){2-5}\cmidrule(lr){6-9}
			h & VAR & LASSO & WLASSO1 & DM p-value & VAR & LASSO & WLASSO1 & DM p-value \\
			\hline
			1 & 2.64 & 2.06 & 2.04 & 0.33 & 2.83 & 2.22 & 2.31 & 0.51 \\ 
			2 & 2.88 & 2.12 & 2.09 & 0.09 & 3.54 & 2.97 & 2.91 & 0.30\\ 
			3 & 2.64 & 2.10 & 2.10 & 0.68 & 3.33 & 2.97 & 2.90 & 0.20\\ 
			4 & 2.40 & 2.10 & 2.10 & 0.77 & 3.12 & 2.90 & 2.84 & 0.17\\ 
			\hline
			& \multicolumn{4}{c}{weekend peak time } & \multicolumn{4}{c|}{weekend off-peak time } \\
			\cmidrule(lr){2-5}\cmidrule(lr){6-9}
			h & VAR & LASSO & WLASSO1 & DM p-value & VAR & LASSO & WLASSO1 & DM p-value \\
			1 & 5.14 & 2.93 & {\bf 2.44} & 0.01 & 3.98 & 2.10 & 1.94 & 0.10 \\ 
			2 & 4.67 & 3.36 & {\bf 3.07} & 0.00 & 5.37 & 2.19 & 2.16 & 0.28 \\ 
			3 & 5.10 & 3.39 & {\bf 3.11} & 0.00 & 7.53 & 2.25 & 2.20 & 0.09 \\ 
			4 & 5.18 & 3.32 & {\bf 3.13} & 0.00 & 9.88 & 2.23 & 2.22 & 0.35 \\ 
			\hline
		\end{tabular}
	}
	\label{table:forecast}
\end{table}

\newpage

\begin{figure}[H]
	\centering
	\begin{tabular}{@{}c@{}}
		\includegraphics[width=1\textwidth]{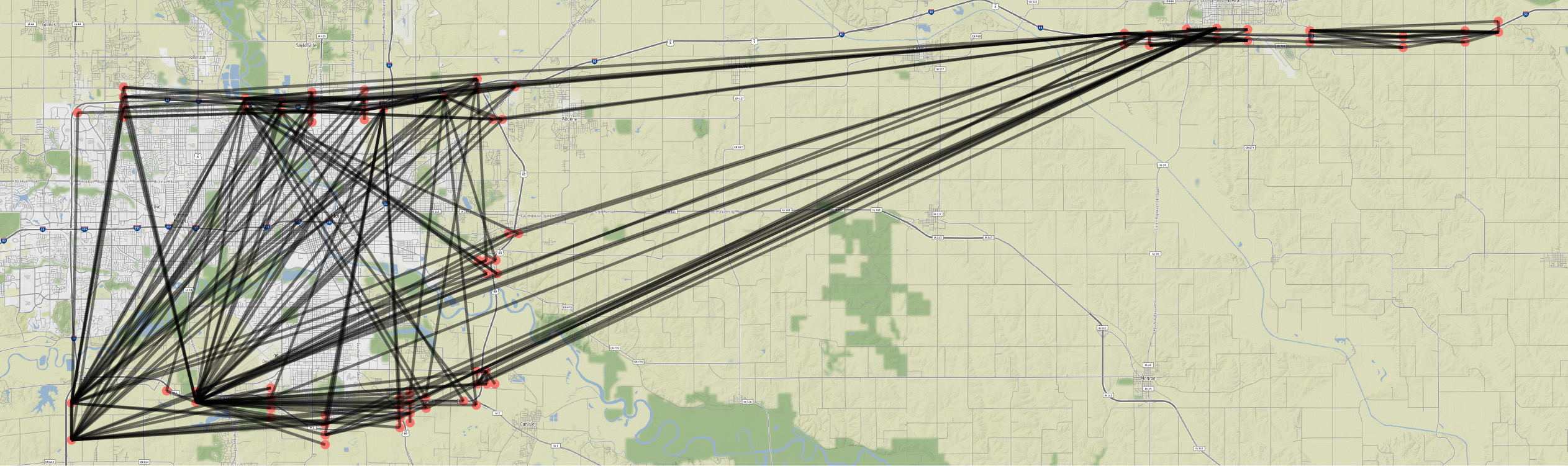}\\
		\includegraphics[width=1\textwidth]{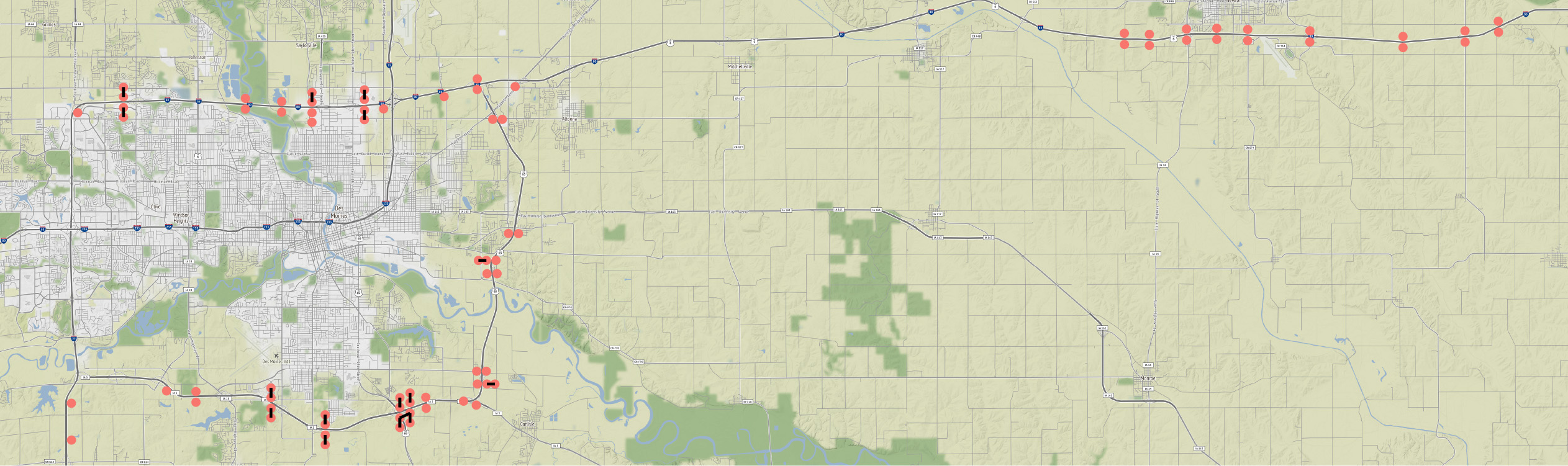} 
	\end{tabular}
	\caption{Estimated network by LASSO (the upper panel) and WLASSO1 (the lower panel) for weekday peak time. If $\hat{\Phi}_{l,ss'}$ is nonzero for at lease one $l$, there will be a connection between site $i$ and site $j$.}
	\label{fig:weekdaypeak}
\end{figure}

\begin{figure}[H]
	\centering
	\begin{tabular}{@{}c@{}}
		\includegraphics[width=1\textwidth]{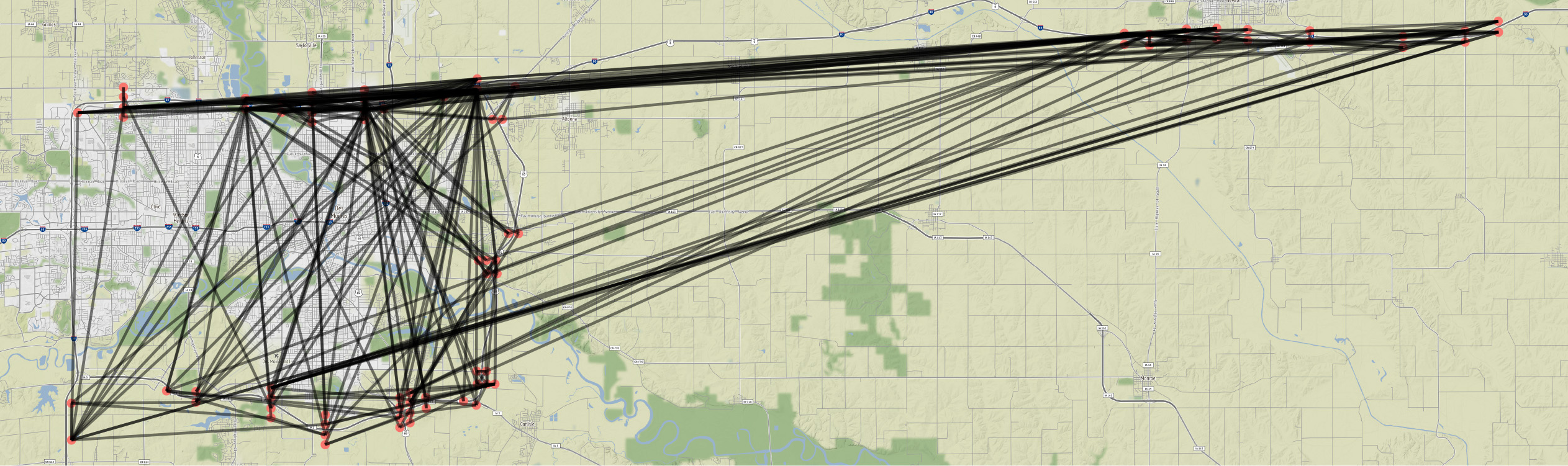}\\
		\includegraphics[width=1\textwidth]{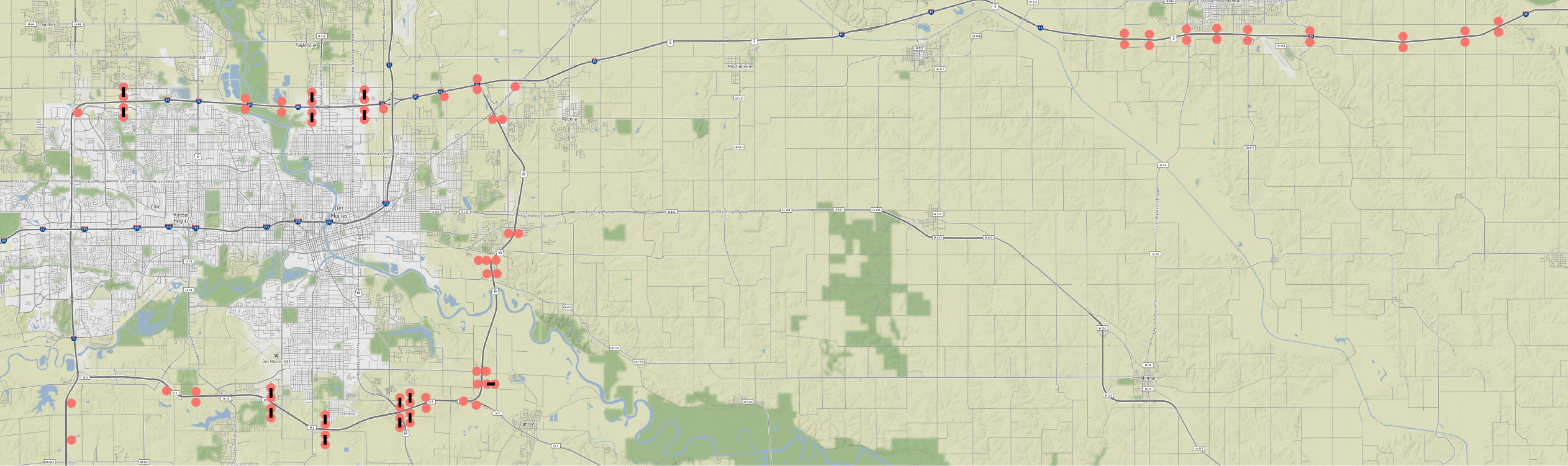} 
	\end{tabular}
	\caption{Estimated network by LASSO (the upper panel) and WLASSO1 (the lower panel) for weekday off-peak time. If $\hat{\Phi}_{l,ss'}$ is nonzero for at lease one $l$, there will be a connection between site $i$ and site $j$.}
	\label{fig:weekdayoff}
\end{figure}

\begin{figure}[H]
	\centering
	\begin{tabular}{@{}c@{}}
		\includegraphics[width=1\textwidth]{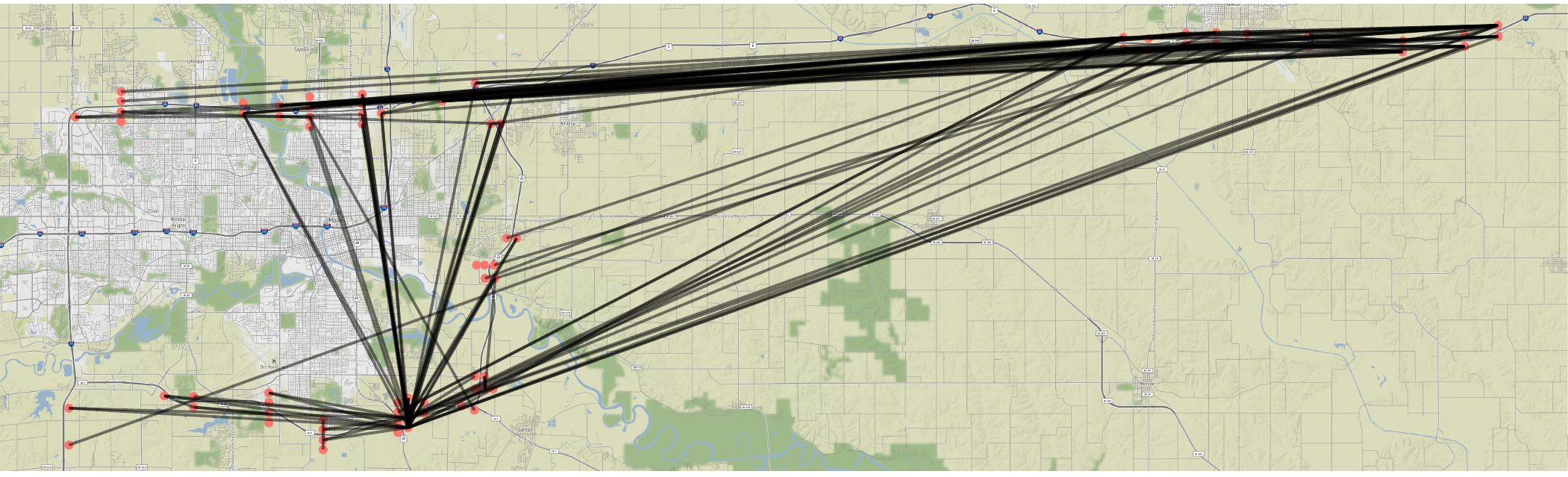}\\
		\includegraphics[width=1\textwidth]{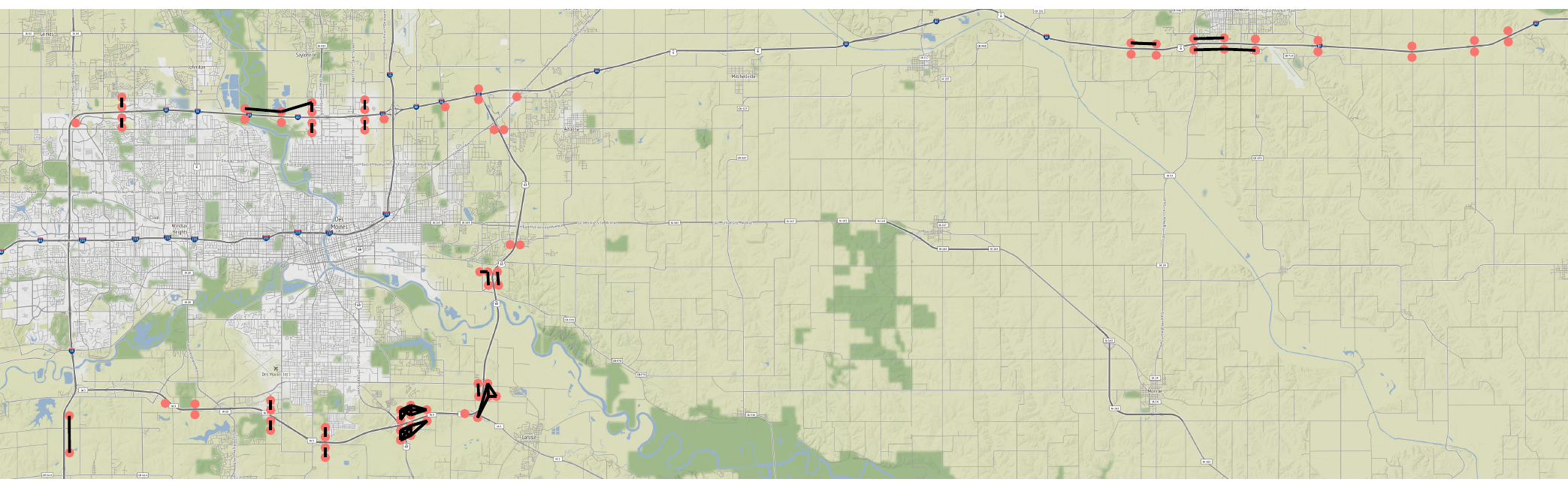} 
	\end{tabular}
	\caption{Estimated network by LASSO (the upper panel) and WLASSO1 (the lower panel) for weekend peak time. For WLASSO1, if $\hat{\Phi}_{l,ss'}$ is nonzero for at lease one $l$, there is a connection between site $i$ and site $j$. For LASSO, since its $\hat{\Phi}$ has too many nonzero, we used a truncated version $\tilde{\Phi}_{ss',l}=\hat{\Phi}_{ss',l}I(|\hat{\Phi}_{ss',l}|\geq0.05)$ to draw the network.}
	\label{fig:weekendpeak}
\end{figure}

\begin{figure}[H]
	\centering
	\begin{tabular}{@{}c@{}}
		\includegraphics[width=1\textwidth]{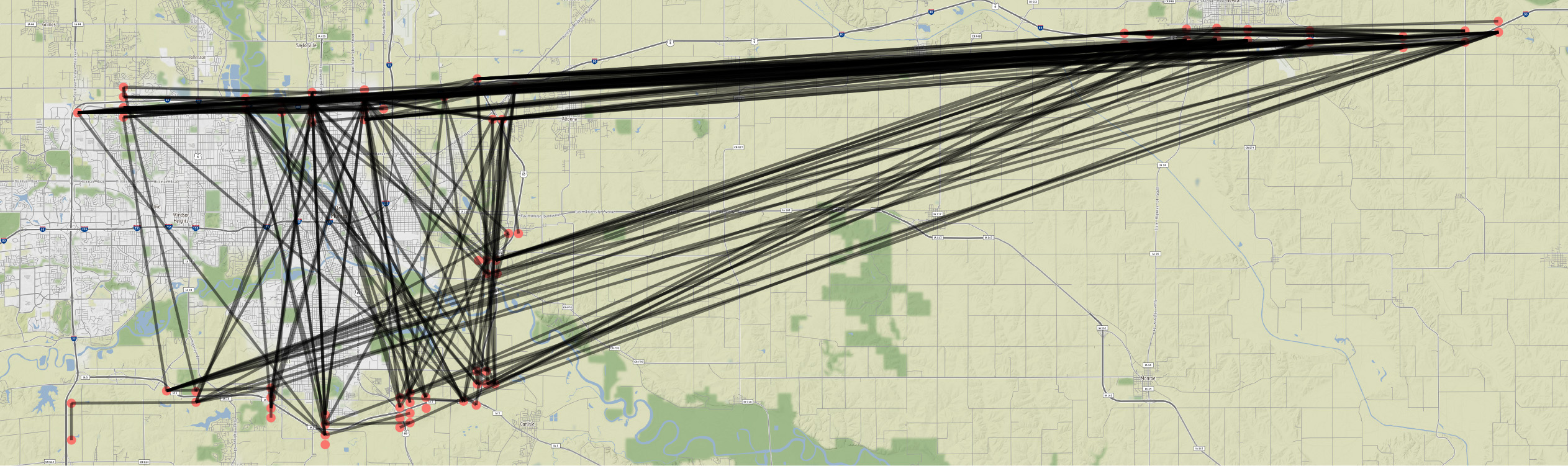}\\
		\includegraphics[width=1\textwidth]{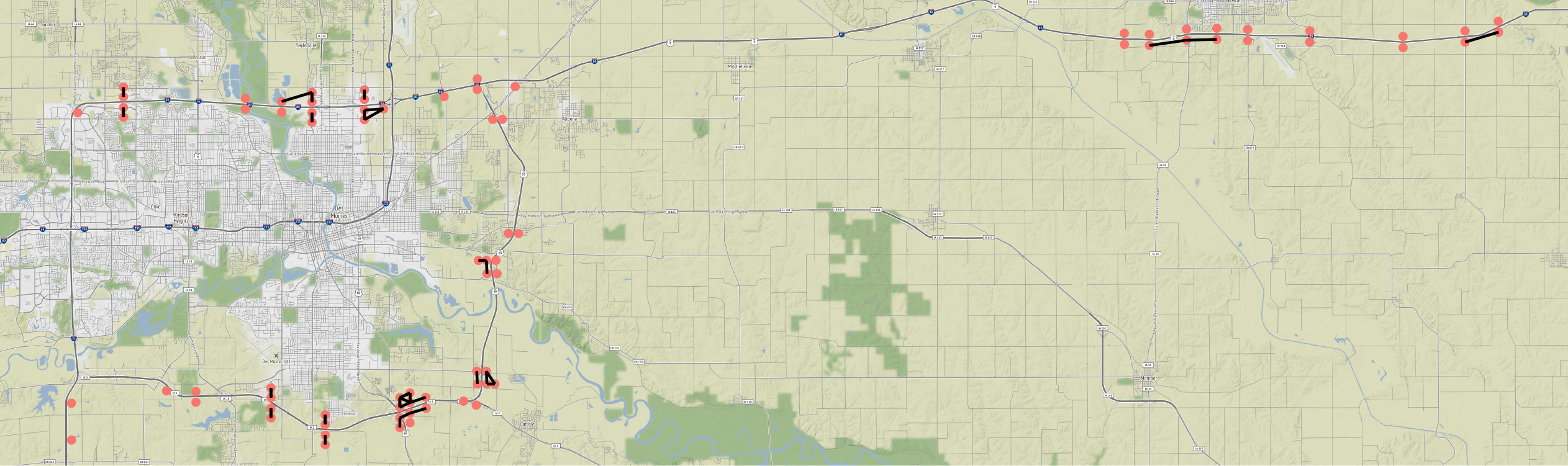} 
	\end{tabular}
	\caption{Estimated network by LASSO (the upper panel) and WLASSO1 (the lower panel) for weekend off-peak time. If $\hat{\Phi}_{l,ss'}$ is nonzero for at lease one $l$, there will be a connection between site $i$ and site $j$.}
	\label{fig:weekendoff}
\end{figure}

\begin{sidewaysfigure}[h]
	\centering
	\includegraphics[scale=0.65]{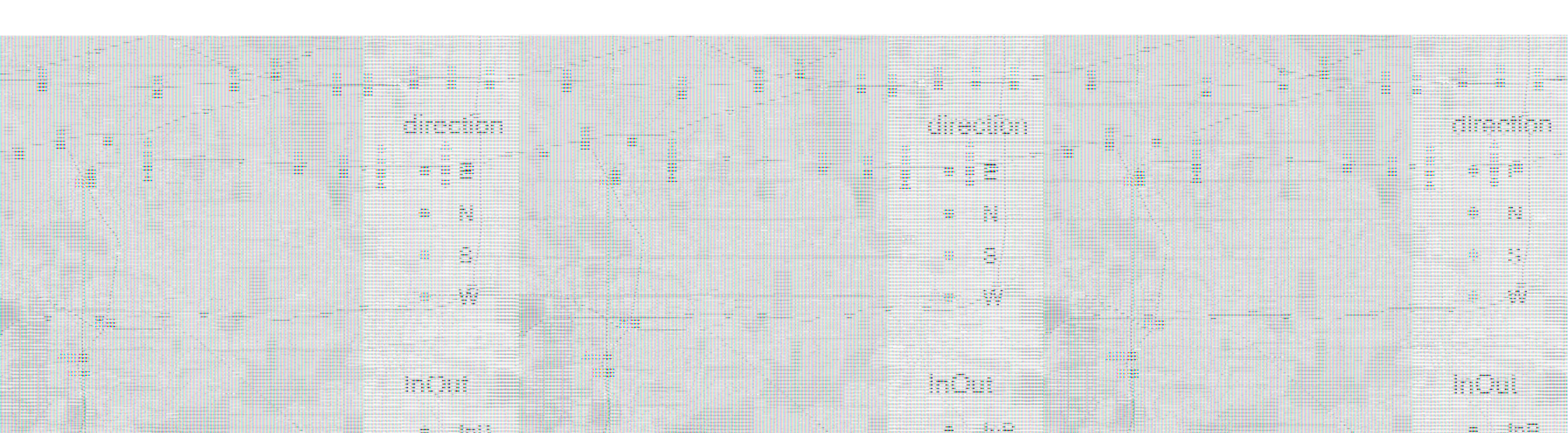}
	\caption{79 sites on highways around Des Moines, Iowa. ``InB" and ``OutB" means entering the highway and exiting the highway, ``Stay" means staying on the highway. ``E", ``W", ``S", ``N" are the directions of traffic flow passing the site.}
	\label{fig:site}
\end{sidewaysfigure}

\end{document}